\documentclass[12pt]{article}
\usepackage[utf8]{inputenc}
\usepackage{graphicx,psfrag,epsf}
\usepackage{booktabs}
\usepackage{textgreek}
\usepackage{threeparttable}
\usepackage{float}
\usepackage{amsmath,amsfonts,amsthm,bm} 
\usepackage{url}
\usepackage{transparent}
\usepackage{adjustbox}
\usepackage{flafter}
\usepackage{lscape}
\usepackage{caption}
\usepackage{subcaption}
\usepackage{graphicx}
\usepackage{geometry}
\usepackage{footnote} 
\makesavenoteenv{tabular} 
\usepackage{verbatim}
\usepackage{natbib}
\usepackage{comment}
\usepackage{rotating}
\usepackage{hyperref}
\usepackage{mdframed}
\usepackage{lipsum}
\usepackage{array}
\newcolumntype{H}{>{\setbox0=\hbox\bgroup}c<{\egroup}@{}}
\usepackage{xcolor}
\usepackage{amsmath}
\usepackage{multirow}
\usepackage{bbm}
\usepackage{algorithmicx}
\usepackage{algorithm,algpseudocode}
\usepackage{longtable}

\newcommand{\blind}{0}
\DeclareCaptionType{myfigure}[Figure]

\begin{document}

\def\spacingset#1{\renewcommand{\baselinestretch}%
{#1}\small\normalsize} \spacingset{1}

\def\spacingset#1{\renewcommand{\baselinestretch}%
{#1}\small\normalsize} \spacingset{1}


\if0\blind
{
  \title{\bf Momentum Informed Inflation-at-Risk\thanks{Tibor Szendrei thanks the ESRC for PhD studentship as well as Heriot-Watt University for institutional support. The authors also thank Geoffrey Hewings for the discussions that shaped the methodology, Mark Schaffer for insights that helped in framing the test, Atanas Christev for advice on motivating the research problem, David Kohns for technical suggestions, and all participants of the PhD conference at Panmure House in 2022 for their invaluable feedback. The usual disclaimers apply.}}
  \author{Tibor Szendrei \footnote{Corresponding author: t.szendrei@niesr.ac.uk.}\\
    Department of Economics, Heriot-Watt University, UK, \& \\
    National Institute of Economic and Social Research, UK. \\
    \\
    Arnab Bhattacharjee\\
    Department of Economics, Heriot-Watt University, UK, \& \\
    National Institute of Economic and Social Research, UK.}
  \maketitle
} \fi

\if1\blind
{
  \bigskip
  \bigskip
  \bigskip
  \begin{center}
    {\LARGE\bf Title}
\end{center}
  \medskip
} \fi


\bigskip
\begin{abstract}
\noindent Growth-at-Risk has recently become a key measure of macroeconomic tail-risk, which has seen it be researched extensively. Surprisingly, the same cannot be said for Inflation-at-Risk where both tails, deflation and high inflation, are of key concern to policymakers, which has seen comparatively much less research. This paper will tackle this gap and provide estimates for Inflation-at-Risk. The key insight of the paper is that inflation is best characterised by a combination of two types of nonlinearities: quantile variation, and conditioning on the momentum of inflation.
\end{abstract}


\noindent%
\vfill

\spacingset{1.45} 


\section{Introduction}
Since the seminal paper of \citet{adrian2019vulnerable}, Growth-at-Risk (GaR) modelling has been a key topic for research and policy. The paper showed how quantile regression of \citet{koenker1978regression} is capable of capturing nonlinearities that are important to the GDP growth distribution. In particular, vulnerability of GDP growth is driven primarily by financial conditions with no evidence of nonlinearity in the coefficient profile for the lag of GDP growth \citep{adrian2019vulnerable}. Another key finding is that the lower quantiles show more variability than the upper quantiles. \citet{figueres2020vulnerable} show similar findings using Euro Area data.

The past years have seen extensive studies on GaR, especially from the perspective of big data. \citet{kohns23hsbqr} find no real difference in variability between upper and lower quantiles. However, before crises episodes the forecast distribution becomes multimodal; see also \citet{mitchell2022constructing} and \citet{adrian2021multimodality}. In relation to financial conditions driving nonlinearities in GaR, \citet{kohns2021decoupling} find evidence of quantile specific sparsity, i.e. different variables are important for different parts of the distribution. \citet{szendrei2023revisiting} also find evidence of quantile specific sparsity for Euro Area data, and show that bank related variables offer better fit than a composite financial indicator. \citet{chen2021quantile} recast quantile regression into a factor framework, and show how quantile factor modelling provides gains in GDP tail estimation. Using a regime switching model, \citet{loria2020understanding} mimic vulnerable growth behaviour without the need for quantile regression. \citet{clark2023tail} also depart from quantile regression, using Bayesian Additive Regression Trees to provide GDP tail estimates. Numerous other important contributions in Growth-at-Risk \citep{choreassessing,galan2020benefits,suarez2022growth,ferrara2022high}, as well as technical- and financial stability reports of central banks, highlight the importance of GDP tail risk.

By contrast, Inflation-at-Risk (IaR) estimation has seen comparatively less interest, but \citet{banerjee2020inflation} and \cite{lopez2022inflation} are important contributions. These papers follow \citet{adrian2019vulnerable} in formulating a macroeconomic relation for inflation tail risk. While the two papers utilise different variables, they both use the Phillips curve as guide in variable selection. In the high-dimensional setting, \citet{korobilis2017quantile} was an early contribution for inflation density forecasts using quantile regression. More recently \citet{lenza2023density} also provided inflation density forecasts. Another important contribution is \citet{pfarrhofer2022modeling}, who utilise time-varying parameter framework to obtain density estimates.

The relatively sparse literature on IaR is particularly surprising given how policymakers are interested in all parts of the inflation density. Forecasted deflation (conditional on covariates) prompts policymakers to take action, just as much as forecasted high inflation (conditional on covariates). This paper aims to fill this gap by providing a method to fit IaR. The key departure from other IaR papers is that we uncover further nonlinearities in the estimates. Specifically, we allow the \textit{momentum} of inflation (i.e. the first difference of inflation) to have an impact on the estimated coefficients. 

The choice of the momentum of inflation as a conditioning variable is rooted in behavioural macroeconomic models about inflation such as \citet{de2011animal}. In such models economic agents have limited cognitive capacity, which prompts them to use biased rules for predicting the future. However, the economic agents adapt and learn from their errors which leads them to change their forecasting strategy. In the realm of monetary policy, the two (biased) strategies are: (1) forecasting inflation with an autoregressive model, and; (2) taking the central bank target at face value. When the momentum of inflation is high, i.e. inflation deviates from the central bank target abruptly, agents are less likely to rely on central bank inflation targets to form their expectations about inflation. As such, we expect different conditional distributions of inflation based on the momentum of inflation.

We find that conditioning on the momentum of inflation has a significant impact on the estimated coefficients. The coefficient profiles of the estimated IaR display variation both across quantiles and inflation momentum. This in turn leads to momentum informed IaR yielding the best measures of in-sample fit, especially at the tails. Considering out-of-sample performance, we also find that our proposed method, combining Conditionally Parametric Quantile Regression (CPQR) \citep{mcmillen2015conditionally} with non-crossing constraints, provides the best forecast performance among the estimators. Furthermore, the CPQR provides considerable gains in forecast performance in longer forecast horizons. Using a Hausman test we can also identify which variables drive the different types of nonlinearities.  We find that lagged inflation has significant quantile variation during periods of extreme momentum in inflation. The quantile profile of GDP growth is mostly present in periods of falling inflation and are less important in periods of growing inflation. In contrast, the quantile varying impact of NFCI is more prominent during periods of falling inflation. These findings highlight that for IaR models, including past values of inflation is not sufficient, one also has to condition on how the economy got there: sudden surges in inflation alter economic agents' expectations differently, which results in different quantile profiles.

The paper is organised as follows. Section~2 utlines the motivation for conditioning on momentum of inflation while Section~3 discusses the methodology, with particular focus on how to condition on the dynamics of inflation within a quantile regression framework using the Conditionally Parametric Quantile Regression. Since the method runs repeated quantile regressions on subsets of the data, we also impose non-crossing constraints to help guide the fitted model. This is followed in Section~4 by discussion of the data and results. In particular, we focus on the coefficient profiles as well as the in-sample fit of the method. We also present out-of-sample fit of the method compared to the traditional quantile regression. Finally, Section~5 collects our conclusions.

\section{Momentum Motivation}
Inflation is a key part of economic agents' lives as it directly impacts their living standards. In periods of high inflation economic agents' consumption responses will depend largely on their wealth profile as highlighted in \citet{lieb2022inflation}. In particular, the authors find that the increased probability to spend when inflation expectations increases is stronger for agents with lower average net worth. This is especially true for agents' with fixed interest rate mortgages. Furthermore, agents' inflation expectations play a crucial role in wage setting dynamics during periods of elevated inflation as shown in \citet{jorda2023inflation}. The authors emphasise that central bank credibility is essential in managing inflation expectations.

While inflation expectations are important for agents' consumption and wage setting decisions, it is also important to note that economic agents rarely form rational inflation expectations. In particular, cognitive biases play a crucial role in how agents form inflation expectations. Economic agents rarely know the ``truth'', and instead understand small parts of the total information available. This leads to agents using heuristics, to form expectations about future outcomes \citep{kahneman2006anomalies,akerlof2010animal,de2011animal}. 

In the model of \citet{de2011animal} these behavioural biases manifest in two strategies for agents inflation expectation formation. In the first strategy the agent believes that the central bank's inflation target will be met and as such will expect inflation to be in line with what the central bank communicates. In the second strategy, the agent forecasts inflation with an autoregressive model. Over time, agents adapt and learn from their forecasting errors, modifying their reliance on these strategies accordingly.

Given this backdrop it is crucial for central banks to accurately forecast inflation for effective monetary policy. Due to the dynamic nature with which inflation expectations are formed it is not surprising that models that rely on assumptions of static parameters have difficulty forecasting inflation accurately. Even models, such as quantile regression, that allow for heterogeneity in parameters showcase a large degree of time variation in the coefficients. In particular, \citet{lopez2022inflation} finds significant time variation in the quantile regression coefficients of a US IaR between 1973 and 2019, highlighting the significant changes inflation has undergone over various regimes in the past 30 years.

The behavioural underpinnings of inflation point towards the potential of these time variations being driven by ``animal spirits'', which are the endogenous waves of optimism and pessimism driven by behavioural biases as described in \citet{de2011animal}. These waves can have a critical role in shaping macroeconomic variables such as inflation. As such, the goal is to propose a empirical strategy that can account for the presence of ``animal spirits''. One promising approach to improve inflation models, such as IaR, is to condition on the momentum of inflation.

By incorporating the rate of change of inflation, rather than just its level, empirical models could reflect the adaptive behaviours of economic agents. When inflation is accelerating or decelerating rapidly, it prompts agents to revise their inflation expectations, potentially away from central bank targets. By incorporating the momentum of inflation into our empirical model, we can potentially account for these psychological and behavioural fluctuations. In essence, sudden changes in inflation can trigger shifts in agents' expectations.

In practice, when inflation momentum is high, it often leads to greater uncertainty and variability in inflation expectations among agents. This is because rapid changes in inflation can undermine the credibility of central bank targets and prompt agents to adjust their expectations more dynamically. Given that the heuristic that different individuals utilise can be diverse, we use quantile regression to capture this heterogeneity. In essence, the quantile varying coefficients, given the degree to which central bank targets are trusted, allow us to capture the heterogeneity of inflation expectation models among the economic agents. In this way the two type of nonlinearities capture different aspects of inflation expectations: momentum conditioning captures the degree of trust in central bank targets, while quantile variation captures the heterogeneity of inflation expectations (given the degree of trust in central bank targets).

Volatility of recent inflation realisations could be an alternative to track the degree of belief in central bank inflation targets. Nevertheless, the momentum of inflation has several key advantages. First, large momentum values will lead to higher volatility, meaning that momentum and volatility often move in conjunction. However, calculating momentum requires only two data points, while volatility needs a longer historical window to assess past fluctuations. Second, relying on a large window size for volatility can potentially dilute the relevance of more recent data, which could pose a problem given the dynamic nature of agents' expectations. Finally, momentum captures the direction of the inflation change, allowing to distinguish between periods of increasing and decreasing inflation, which is not possible with volatility of inflation. The ability to distinguish the direction of inflation change is crucial as individuals are likely to react differently to rapidly increasing and decreasing inflation. These points make momentum a better candidate for conditioning variable to model adaptive behaviours in inflation dynamics.

Combining quantile regression and momentum conditioning aligns with the behavioural macroeconomics perspective. This empirical approach is an attempt at recognising that agents' reactions to changing economic conditions are not purely rational but influenced by psychological factors and cognitive biases. As a result, incorporating momentum conditioning with quantile regression helps capture a spectrum of agents' adaptive responses and offers a more comprehensive way of modelling inflation dynamics.



\section{Methodology}

The core idea of macroeconomic at-risk models is to use quantile regression of \citet{koenker1978regression} to capture the nonlinear impact of financial conditions on macroeconomic variables. Formally, this is done by minimising the weighted least absolute deviation:
\begin{equation}\label{eq:QR5}
\begin{split}
    \hat{\beta}_\tau&=\underset{\beta_\tau}{argmin}\sum^{T-h}_{t=1}\rho_\tau(y_{t+h}-x_t^T\beta_\tau)\\
    \rho_\tau(u)&=u(\tau-I(u<0)).
\end{split}
\end{equation}

By including an intercept and several other covariates in equation (\ref{eq:QR5}), one is able to characterise the conditional distribution of the variable of interest. Key covariates included in most macroeconomic at-risk models are the lag of the variable and a variable capturing financial conditions ($F_t$). 
Additional covariates can be included as well which leads to an at-risk model of the following form:
\begin{equation} \label{eq:XaRmodels}
    \hat{\mathcal{Q}}_\tau(y_{t+h}|x_t)=\beta_{0,\tau} + \beta_{1,\tau} y_t + \sum^{K-1}_{k=2} \beta_{k,\tau}x_{k,t}+\beta_{K,\tau} F_t.
\end{equation}



\newpage
The additional nonlinearity we will introduce is momentum-thresholding/conditioning. The central idea stems from \citet{enders1998unit} and later \citet{enders2001cointegration}, that rather than the level of some variable, the first difference, i.e., the momentum, of the variable acts as a thresholding variable. 
To introduce additional nonlinearities for IaR, there are two candidate approaches: Threshold Quantile Autoregression (TQAR) \citep{galvao2011threshold}; and Conditionally Parametric Quantile Regression (CPQR) \citep{mcmillen2015conditionally}.\footnote{We note that tail estimates during large changes in inflation are difficult to capture since such periods are rare which leads to less data being available. This also justifies non-crossing constraints.}

Using the TQAR raises consideration of several critical questions. First, should we model a joint threshold shared across quantiles, or quantile specific thresholds? While quantile specific thresholds allow more flexibility, it is not clear whether all thresholds can be identified simultaneously. Second, do we allow for the number of regimes to be quantile specific or shared across the distribution? Allowing for different number of regimes per quantile is appealing but increases model complexity. Finally, how many thresholds do we allow for each quantile? \textit{Ex ante} it is difficult to ascertain the number of regimes the different quantiles might have. Furthermore, jointly identifying the number of thresholds for each quantile is not a trivial task. By contrast, applying CPQR is less ambiguous needing only a grid of values of the conditioning variable and a bandwidth.\footnote{Technically, a kernel is also required, but the choice of kernel has limited effect on the final estimates, as compared to a change in the bandwidth parameter.} Therefore, we choose to use the CPQR to condition on the momentum of inflation.

\subsection{Conditionally Parametric Quantile Regression (CPQR)}

The key problem is that the fully parametric model of TQAR is not known with certainty. Then, some structure can be used to circumvent the complexity of a fully parametric model. Conditionally parametric models, as discussed in \citet{cleveland1994coplots}, can be used in such situations. In essence, the method applies a parametric model for a set of covariates $x$, conditional on a set of variables $z$. This way a nonparametric model $y_t=f(x_t,z_t)+u_t$ can be restricted to the form $y_t=x_t\beta(z_t)+u_t$. Importantly, the set of conditioning variables $z$ can be a subset of $x$. As such, the conditionally parametric model can be viewed as a nonparametric version of a threshold model when the threshold structure is unclear or difficult to pin down.

This type of conditionally parametric model is popular in spatial contexts, where it is commonly called geographically weighted regression \citep{mcmillen1996one}, and was adapted to quantile regression by \citet{mcmillen2015conditionally}. Then, conditioning on $z$, the conditionally parametric quantile regression estimator is given by:
\begin{equation}\label{eq:CPQR}
\begin{split}
    \hat{\beta}_\tau&=\underset{\beta_\tau}{argmin}\sum^{T-h}_{t=1}w_t(z)\rho_\tau(y_{t+h}-x_t^T\beta_\tau)\\
    w_t(z)&=K\Big(\frac{z_t-z}{b_h}\Big),
\end{split}
\end{equation}
where $z$ is the target for the weighted quantile regression, $K(u)$ is some kernel weight function, and $b_h$ is the bandwidth. In essence, equation (\ref{eq:CPQR}) is just a locally weighted quantile regression, where the weights are regulated by the distance from $z$ \citep{chaudhuri1991nonparametric, yu1998local}.

Equation (\ref{eq:CPQR}) is simply a weighted version of (\ref{eq:QR5}), where the observations weights are determined by a kernel function. The novelty here is that we apply this estimator to a non-spatial setting. In particular we will create a conditional IaR model, where 
the choice for conditioning variable is the lag of momentum of inflation, $\Delta y_{t-1}$. This way the model will provide quantile regression estimates conditional on the \textit{momentum} of inflation. Weights need to be assigned to the observations, depending on the value of the change in inflation, and determined by the kernel and the bandwidth of equation (\ref{eq:CPQR}). Following \citet{mcmillen2015conditionally} the tri-cube kernel is chosen with a window size of $d_\tau$, i.e. the nearest $d_\tau$ observations to the specific inflation momentum are used. Formally:
\begin{equation}\label{eq:kernel}
    w_t(z)=I(d_t \leq d_\tau)\frac{\Big( 1-\Big( \frac{d_t}{d_\tau} \Big)^3 \Big)^3}{d_\tau}
\end{equation}

While there are several other choices of kernels available, the choice of kernel tends to have a limited impact on results \citep{mcmillen2012quantile,mcmillen2015conditionally}. The bandwidth on the other hand has a far greater influence on the estimation. To make the bandwidth choice principled, $d_\tau$ is chosen by cross-validation as suggested by \citet{racine2004nonparametric}. The cross-validation exercise takes the final 10\% of the observations and does one-step at a time forecast for a grid of bandwidths. For each forecast period, the bandwidth that leads to the lowest sum of weighted deviations is recorded. The final bandwidth, $d_\tau$, is the bandwidth that yields the lowest average weighted residual. The grid of bandwidths used for this search is each $5^{th}$ quantile starting from the $10^{th}$ up until the $90^{th}$. The reason for the bandwidths being limited to be larger than the $10^{th}$ quantile is to ensure that there are sufficient number of observations to run the $\tau$ quantile regressions. The described cross-validation method helps with automatically adjusting the bandwidth as the available data become bigger. The advantage of using quantiles as bandwidths is that it is easy to control and interpret the impact it has on the estimation: larger bandwidth leads to more data being considered and smoother quantiles \citep{mcmillen2015conditionally}.

\begin{figure}[]
    \centering
    \includegraphics[width=\textwidth]{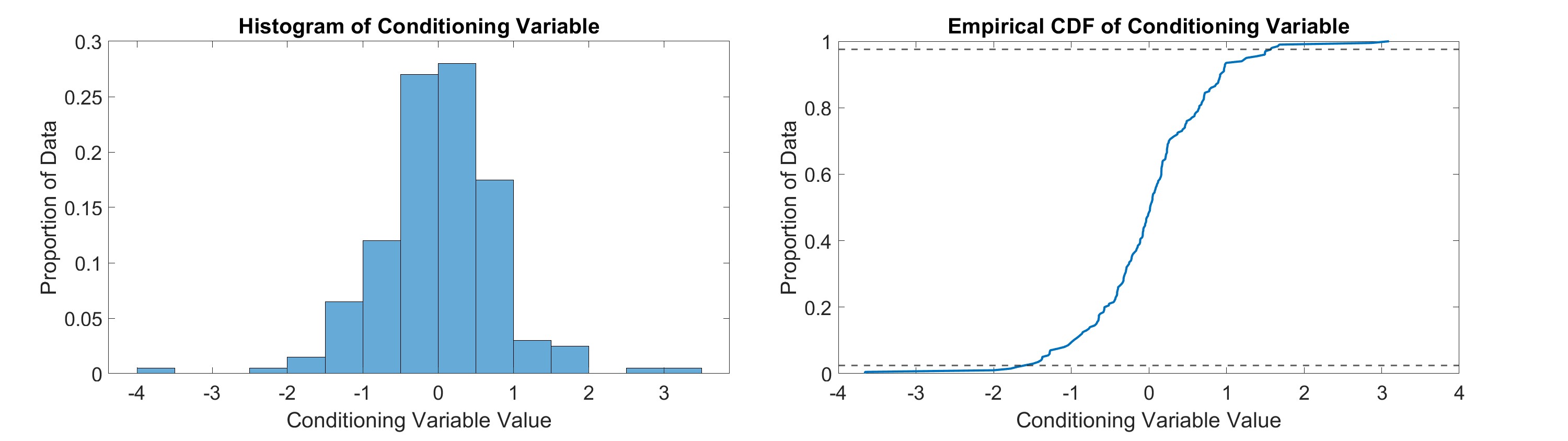}
    \caption{Histogram and CDF of $\Delta y_{t-1}$}
    \label{fig:MomentumHist}
\end{figure}

\newpage
The second important consideration for the CPQR is the grid of conditioning variables. Figure (\ref{fig:MomentumHist}) shows the histogram and CDF of the momentum of inflation. One can see that over 95\% of the values are between -2 and 2. The 95\% band is shown in the CDF with horizontal dashed lines. For this reason we consider a grid of conditioning values between -2 and 2, with the grid fineness set to 0.2. This ensures that there is sufficient data at each considered conditioning value.

When comparing CPQR and normal QR on the IaR model of equation (\ref{eq:XaRmodels}), it is important to ensure that the addition of nonlinearities is driving the difference in fit. To this end, we also estimate normal quantile regression with the momentum of inflation added as an additional covariate. Note that this is equivalent to estimating equation (\ref{eq:XaRmodels}) with two lags \citep{wolters2015changing}. Hence, we refer to this model as the QAR(2) model, in line with the terminology used in \citet{koenker2006quantile}.

Note that it is possible to have a separate bandwidth parameter for each quantile considered; since more extreme quantiles are determined by fewer observations we expect to have larger bandwidths for these quantiles. Nevertheless, in this paper we choose to select only a single bandwidth parameter and instead shrink unnecessary quantile variation with the use of non-crossing constraints, as in \citet{szendrei2023fused}. This also helps in tackling the problem of quantile crossing occurring due to finite sample sizes and addresses limited data issues at the tails.

\newpage
\subsection{Non-crossing Constraints}

Non-crossing constraints incorporated into quantile regression is a way to ensure that the estimated quantiles remain monotonically increasing. The alternative, called quantile crossing, can occur due to insufficient data or model misspecification \citep{koenker2005}. Importantly, non-crossing constraints not only ensure monotonically increasing quantiles but also helps in limiting the `jaggedness' in the quantile profie \citep{szendrei2023fused}. This latter point is particularly important in the current application, since limited data availability often leads to such non-smooth coefficient profiles. As discussed in \citet{bondell2010noncrossing}, non-crossing constraints can be implemented in quantile regression via inequality constraints:

\begin{equation} \label{eq:NCQR5}
    \begin{split}
        \hat{\beta}_\tau&=\underset{\beta_\tau}{argmin}\sum^Q_{q=1}\sum^{T-h}_{t=1}w_t(z)\rho_{\tau_q}(y_{t+h}-x_t^T\beta_{\tau_q})\\
        &s.t.~x^T\beta_{\tau_q} \geq x^T\beta_{\tau_{q-1}},
    \end{split}
\end{equation}
where $\hat{\beta}_\tau=\{\hat{\beta}_{\tau_1}, \hat{\beta}_{\tau_2},...,\hat{\beta}_{\tau_Q}\}$, i.e. the set of parameters for all estimated quantiles. While conceptually simple, the estimation of equation (\ref{eq:NCQR5}) would require the inclusion of $(T-h) \times (Q-1)$ constraints. To reduce the number of constraints we follow \citet{bondell2010noncrossing}, and impose a single constraint across time which represents the worst case scenario possible in our data, denoted as $x_{WC}$.\footnote{For more information on the `worst-case' scenarios for ensuring non-crossing constraints, please refer to \citet{bondell2010noncrossing} or \citet{szendrei2023fused}.} Note that we need a a set of non-crossing constraints for each conditioned value. Since the way we calculate fitted values is by taking the closest grid point to the realised $\Delta y_{t-1}$, we only need to consider imposing non-crossing for data points near the grid point. As such we construct a variable, denoted as $i_{t}(z)$, which takes on a value of 1 if the data falls between the halfway points of the current conditioning values, $z$, adjacent grid points, and 0 elsewhere. This way we ensure non-crossing locally, using only the data points that would be used when calculating the fitted quantiles, conditional on the momentum of inflation. With this in mind our Conditionally Parametric Quantile Regression model is defined as:
\begin{equation} \label{eq:CPQRNC}
    \begin{split}
        \hat{\beta}_\tau&=\underset{\beta_\tau}{argmin}\sum^Q_{q=1}\sum^{T-h}_{t=1}w_t(z)\rho_{\tau_q}(y_{t+h}-x_t^T\beta_{\tau_q})\\
        &s.t.~I(i_{t}(z)>0)x_{WC}^T\beta_{\tau_q} \geq I(i_{t}(z)>0)x_{WC}^T\beta_{\tau_{q-1}}.
    \end{split}
\end{equation}

It is important to note that in equation (\ref{eq:CPQRNC}) the worst case scenario is defined only on the observations that are used for fitting the quantiles as identified by $i_t(z)$, i.e. the constraint is conditional on the momentum of inflation as well, but only locally. While it is possible to impose constraints that are based on all data used in estimation, doing so would be too stringent and would lead to overshrinking quantile variation. In essence, if the interest is only to ensure non-crossing quantiles, the constrained quantile regression should have the same limiting distribution as the classical quantile regression (conditional on $w_t(z)$) \citep{bondell2010noncrossing}. This is the case only if we condition both the objective function and the constraint.

The novelty of equation (\ref{eq:CPQRNC}) is that it estimates conditionally parametric quantiles in such a way that they do not cross in-sample. Additionally, the paper extends the conditionally parametric framework beyond the spatial domain. In particular, we incorporate momentum conditioning to quantile regression to account for the dynamic behaviour of agents' inflation expectations.

\subsection{Hausman test}

The CPQR is capable of modelling various different types of processes since it captures two types of nonlinearities: 1) nonlinearities that arise from momentum in inflation; and 2) nonlinearities that stem from quantile variation given the value of $\Delta y_{t-1}$. We argue that the distinction between these two nonlinearities is that momentum conditioning measures agents' trust in central bank targets, and as such how effective forward guidance is, while quantile regression captures the heterogeneity of agents' inflation expectations given their belief in central bank targets. Given these two sources of nonlinearities, in some situations it might be imperative to pinpoint which type of nonlinearity is driving the process. In particular, identifying when moment conditioning and quantile variation jointly drives the inflation process allows policy-makers to gauge how effective forward guidance will be: since quantile regression is employed to capture agents' heterogeneity in inflation expectations, it is expected that quantile variation will be more important when there is less trust in central banks inflation targets, i.e. in situations of extreme conditioning values. In such situations forward guidance will be less effective.

It may also be informative for policy-makers to know which variables are driving the nonlinearity. Knowing which variables contribute to the nonlinearities is important as it can identify the channel that has the most impact on future inflation. For example, the importance of a variable tracking the real sector can signify increased heterogeneity in wage-setting behaviour among agents, while the importance of a variable tracking the pass-through channel can indicate that inflation is due to supply-side disturbances. As such, we expect more variables to exhibit prominent quantile variation as trust in central bank targets is lost, i.e. for more extreme values of momentum inflation. A better understanding of what variable drives the nonlinearity can also highlight what economic agents' base their future inflation expectations on, which the central bank can use to tailor its communication to maintain agents' trust.

To achieve these goals, we note that the QAR(2) is an efficient version of the CPQR along the momentum dimension, i.e., where one assumes that all nonlinearities are driven by quantile variation. As such, we can compare the coefficients of the QAR(2) and CPQR to determine where conditioning on the momentum of inflation is significantly different. Furthermore, notice that the composite quantile regression (CQR)\footnote{The CQR is an estimator where the coefficients for every covariate (except the intercept) are assumed constant across the quantiles} of \citet{koenker1984note} and \citet{zou2008composite} is also an efficient version of the CPQR, but along the quantile dimension, i.e. where one assumes that all nonlinerities are driven by momentum conditioning. As such one can recast the question of identifying drivers of nonlinearity as a Hausman test of \citet{hausman1978specification}. 

The Hausman test is used to evaluate the consistency of an estimator compared to an alternative, more efficient estimator. In essence the test determines whether the simpler model (assuming no significant differences) is adequate at capturing variability or if the more complex model is necessary. Given, that the CPQR has two efficient versions along the two nonlinearities, we will propose 2 Hausman tests. In the first test we check how much momentum conditioning is important where the common variables of QAR(2) provide the efficient estimator, and the CPQR is the consistent estimator. In the second test we check whether quantile variation is important. Here the common variables of the CQR provide the efficient estimator for the CPQR.


The two tests for checking the presence of nonlinearity are fundamentally Wald tests, which evaluates the null hypothesis that the difference between the efficient and consistent estimators is zero. For this comparison we use the differences in coefficients along with a measure of variability for said differences. As such, the test also requires the variance associated with said coefficients. For this purpose, we use a simple block bootstrap routine to obtain standard errors for the coefficients.\footnote{When estimating the CQR, we select a different bandwidth using the same method used for bandwidth selection of the CPQR.} To ease computational burden we hold the bandwidth needed for CQR and CPQR fixed across the different bootstrap samples.\footnote{Given the findings of \citet{bates2024cross}, the bootstrap-based confidence intervals for the estimators may be too narrow. Therefore, we will rely on the Hausman test results primarily in cases where the coefficient profiles suggest potential nonlinearities.}

Once the coefficients and standard errors of the different estimators are obtained we apply the following computation:
\begin{equation}
    H=(vec(\iota_R \otimes \beta_E)-\beta_C)'\Big(Var(\beta_C) - vec(\iota_R \otimes Var(\beta_E)) \Big)^{\dag}(vec(\iota_R \otimes\beta_E)-\beta_C),
\end{equation}
where $\beta_E$ is the efficient coefficient estimator (either CQR or QAR(2)), $\beta_C$ is the consistent estimator (CPQR), $Var(\cdot)$ is the variance operator, and $^{\dag}$ denotes the Moore–Penrose pseudoinverse.\footnote{The Moore–Penrose pseudoinverse is used as it can even handle cases where the variance matrix is not be invertible.} To ensure that the vector of coefficients and variances are of the same dimension, we take the Kronecker product of $\iota_R$ and $\beta_E$, where $R$ is the number of times the vector needs to be repeated so that it is of the same dimension as $\beta_C$. The reason we need to take the vector operator on the kronecker product is because the two types of nonlinearities are stored in different dimensions. As such, $\iota_R$ can be either a row vector (for CQR) or a column vector (for QR) of ones. In the same vein, $\beta_E$ is either a row vector (for QR) or a column vector (for CQR). With the use of the vectorise operator both sets of vectors are of the same dimension, while structuring the coefficients and $\iota_R$ as either row or column vectors ensures that the right coefficients of the different models are compared with each other regardless of which type of nonlinearity we assume is fixed.

\section{Data and Results}
\subsection{Data}

\begin{table}[]
\centering
\begin{tabular}{p{0.1\textwidth}|p{0.3\textwidth}|p{0.5\textwidth}}
\hline
\multicolumn{1}{c|}{\textbf{Variable}} & \multicolumn{1}{c|}{\textbf{FRED code}} & \multicolumn{1}{c}{\textbf{Variable Description}} \\ \hline \hline
Inflation & CPIAUCSL\_PC1 & Consumer Price Index for All Urban Consumers: All Items in U.S. City Average, Percent Change from Year Ago, Quarterly, Seasonally Adjusted \\ \hline
GDP growth & A191RL1Q225SBEA & Real Gross Domestic Product, Percent Change from Preceding Period, Quarterly, Seasonally Adjusted Annual Rate \\ \hline
Import Prices & B021RG3Q086SBEA\_PC1 & Imports of goods and services (chain-type price index), Percent Change from Year Ago, Quarterly, Seasonally Adjusted \\ \hline
Financial Conditions & NFCI & Chicago Fed National Financial Conditions Index, Index, Quarterly, Not Seasonally Adjusted \\ \hline
\end{tabular}
\caption{Variables of the model} \label{tab:VaribTab}
\end{table}

Our empirical work is based on quarterly macroeconomic data for the US over the period 1973Q1 to 2022Q4. This period covers several different inflation ``regimes'' such as the high inflation era of the 1980s, the great moderation, the great financial crisis of 2008, and the inflation following the COVID-19 global pandemic. It is not surprising that \citet{lopez2022inflation} find extensive time-variation in the estimated quantile coefficients across the diverse inflation ``regimes''. Inflation is measured by the year on year growth rate of the quarterly Consumer Price Index. As conditioning variable, we use the quarterly lag of momentum of inflation, where momentum is measured by the first difference of inflation. As discussed above, figure (\ref{fig:MomentumHist}) shows the distribution of the momentum of inflation.

The proposed method, which combines CPQR with non-crossing constraints, is used to estimate a model for Inflation-at-Risk (IaR) explained by four covariates. The first two covariate is lagged inflation (level), to track persistence in inflation, and the National Financial Conditions Index (NFCI), to allow for the financial sector to have an impact on future inflation. The NFCI is a common choice to track the financial sector in US at-risk models \citep{adrian2019vulnerable,lopez2022inflation}. The NFCI is an index combining information from 105 measures of financial activity \citep{brave2018diagnosing}. When the NFCI is positive, then financial conditions are tighter than on average which can be interpreted as financial stress being present. A variables own lag and a measure of financial conditions form the base of many macroeconomic at-risk models described by equation (\ref{eq:XaRmodels}). 

For the additional covariates we follow \citet{banerjee2020inflation} in including real GDP growth to allow for output to have an impact on future inflation. Rather than including exchange rate and oil price changes as covariates, we follow \citet{lopez2022inflation} and instead include changes in relative import prices. This measure was proposed by \citet{blanchard2015inflation} to capture pass-through of nominal exchange rates and oil prices into inflation, and is particularly useful when inflation is driven by changes in global commodity prices. The raw variables and their respective FRED codes are presented in table (\ref{tab:VaribTab}).


We consider two forecast horizons: one quarter ahead ($h=1$) and one year ahead ($h=4$). For the longer forecast horizon ($h=4)$ we follow \citet{adrian2019vulnerable} and take the four quarter average of inflation.


\subsection{Coefficient profiles}

\begin{figure}
    \centering
    \includegraphics[width=\textwidth]{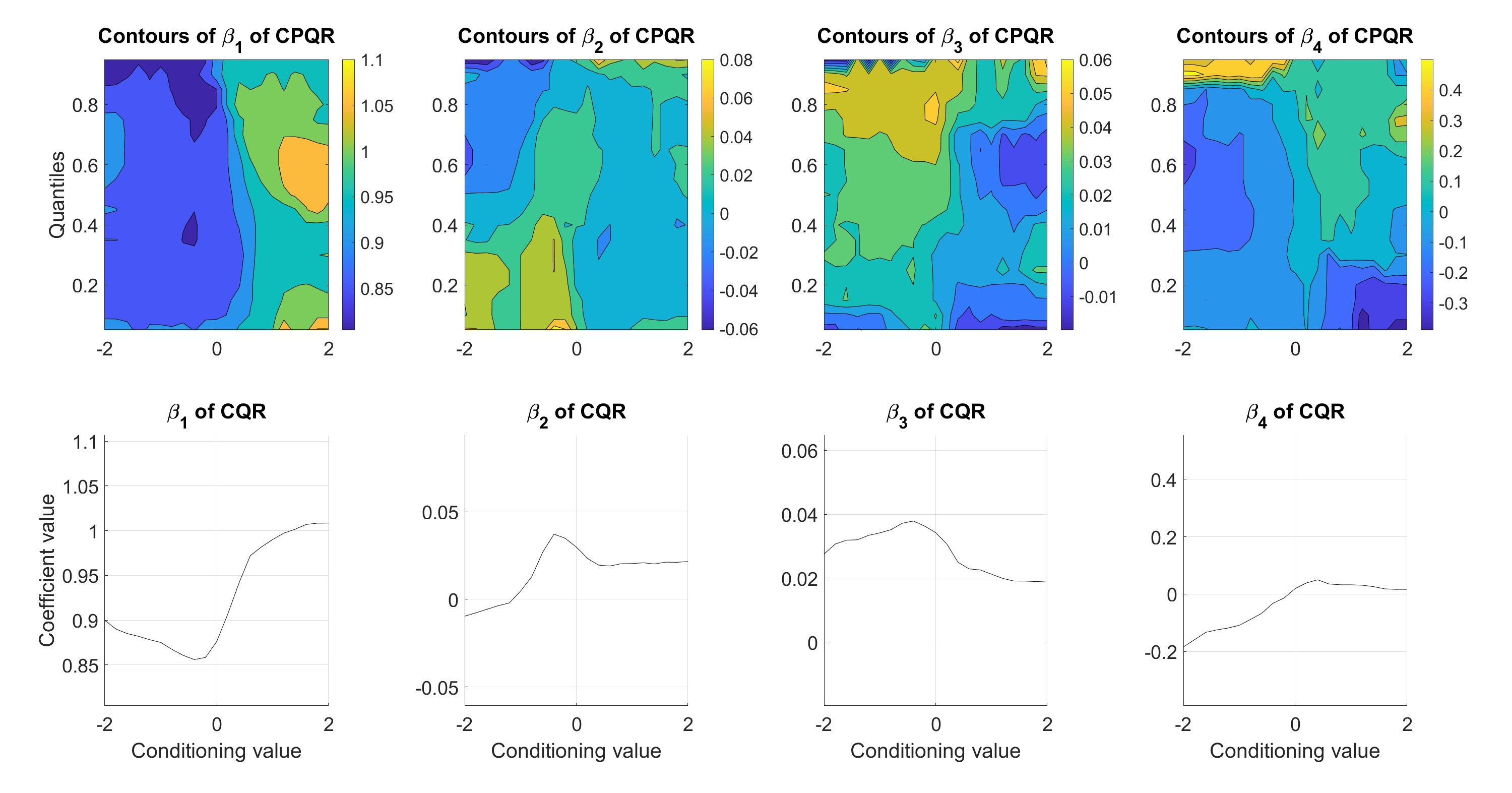}
    \caption{Coefficients of CPQR for $h=1$}
    \label{fig:CPQRCoeff_h1}
\end{figure}

\begin{figure}
    \centering
    \includegraphics[width=\textwidth]{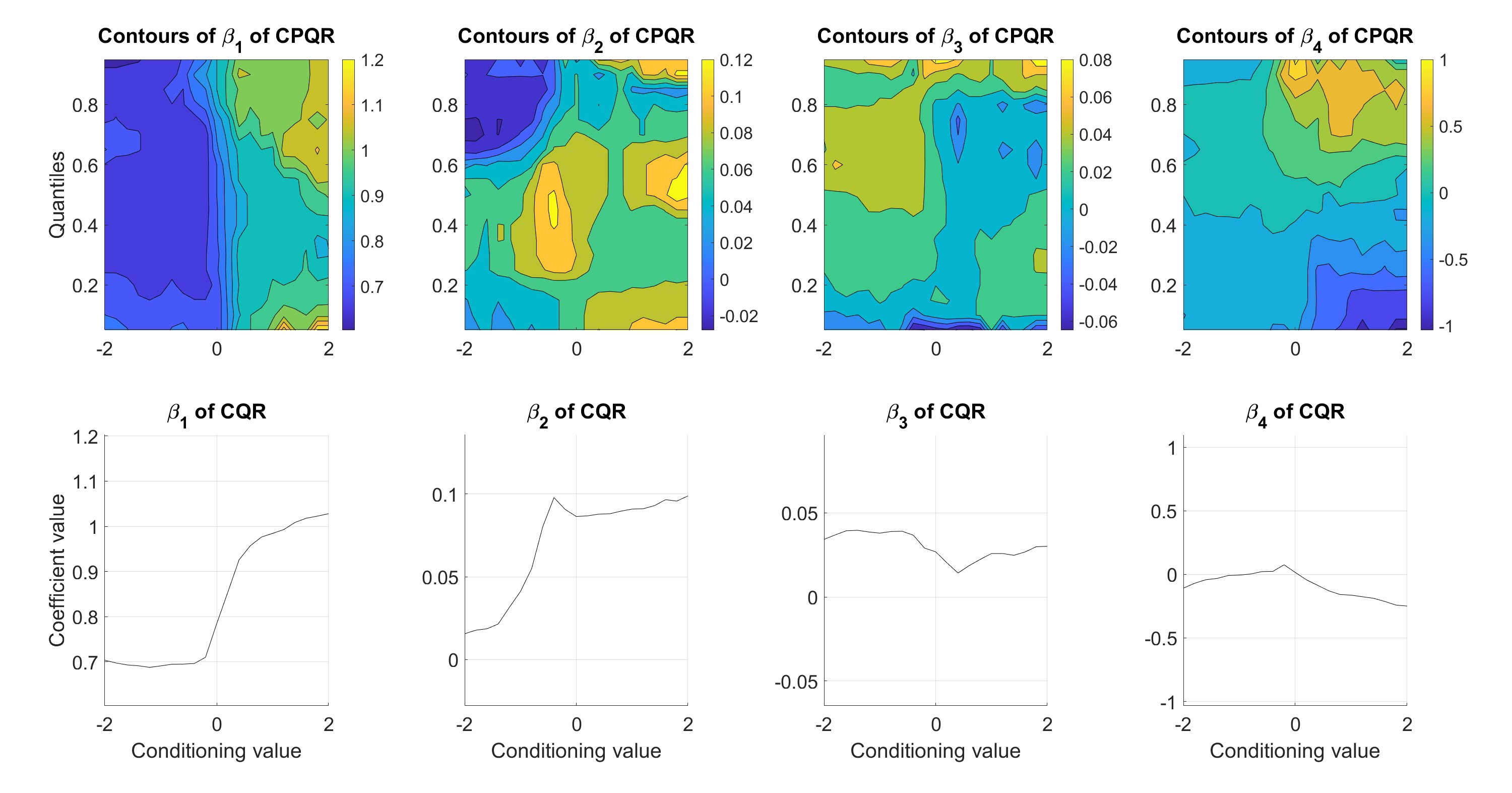}
    \caption{Coefficients of CPQR for $h=4$}
    \label{fig:CPQRCoeff_h4}
\end{figure}

Coefficient profiles for one quarter ahead IaR and one year ahead IaR are shown in figures (\ref{fig:CPQRCoeff_h1}) and (\ref{fig:CPQRCoeff_h4}) respectively. Due to the 3 dimensional nature of the coefficients, we show contour plots of the coefficient profiles in the top row of each figure. In addition, we also provide coefficients of the CQR (bottom row). This provides a visual representation of nonlinearities stemming from quantile variation and from conditioning on the momentum of inflation.

In these figures, $\beta_1$ denotes the coefficient on lagged inflation, $\beta_2$ refers to (lagged) GDP growth, $\beta_3$ corresponds to (lagged) change in relative import prices, and $\beta_4$ to (lagged) NFCI. From the contour plots, it is clear that there is significant variation across both dimensions. Interestingly, the general shape of almost all of the coefficients varying by conditioning state (as captured by CQR) are very similar across the forecast horizons considered: $h=1$ and $h=4$. The only exception is the NFCI, which shows clear difference in the coefficient profile of the CQR between the two forecast horizons. The same is not true for quantile profiles, where there are distinct differences between the two forecast horizons for all covariates.

Considering the covariates individually, we replicate a previous finding that the effect of inflation persistence ($\beta_1$) exhibits quantile variation \citep{wolters2015changing}, but our new result is that it varies more by the momentum of inflation. In particular, periods when inflation is either stable or decreasing are associated with far less inflation persistence than periods with rising inflation. Also, periods with large increase in inflation present (near) unit-root character where the coefficient is close to 1. This is true for both horizons considered, and adds to the finding in the literature that upper quantiles of inflation are likely to be more persistent. Importantly, persistence in inflation varies not just by quantiles, but also with the momentum of inflation. The fact that higher changes in inflation is followed by more persistence is likely driven by agent expectations. In essence when economic agents experience higher momentum in inflation, they are likely to expect further high inflation. Furthermore, the figures also indicate that periods with extreme momentum in inflation have higher quantile variation. This reflects more variation in agents expectations in increasing inflationary periods.

By contrast, the coefficient on real GDP growth ($\beta_2$) shows greater quantile variation than inflation for both forecast horizons. In particular, quantile variation is particularly high during periods of falling inflation (or negative momentum). Similar to \citet{banerjee2020inflation}, the contour plots reveal an overall decreasing profile across quantiles. This suggests that, following a fall in inflation, GDP growth decreases the conditional variance of inflation. However, the quantile profile is less pronounced especially for the shorter forecast horizon following a quarter when inflation has risen. This adds a caveat to the finding of \citet{banerjee2020inflation}, namely that variance reduction impact of GDP growth only follows a period of falling inflation.

For the coefficient of relative import price inflation ($\beta_3$), we observe greater quantile variation for $h=1$. In particular, for $h=4$, the quantile profile is most pronounced for periods where inflation is stable, when the momentum of inflation is near 0. For other conditioned values the results mirror  \citet{lopez2022inflation}, i.e. quantile variation is largely driven by the upper quantiles. By contrast, for $h=1$, there is significantly more quantile variation regardless of the momentum of inflation.

Finally, the coefficient for NFCI ($\beta_4$) also shows distinctly different patterns for $h=1$ and $h=4$. For the short forecast horizon, there is significant quantile variation when the momentum of inflation is \textit{not} near 0. At the longer forecast horizon, the quantile variation of NFCI is more pronounced when inflation is increasing. In particular, we can see that the coefficient profile shows increasing quantile profile of the NFCI when $h=4$, which highlights how financial conditions increase the variance of inflation during periods of rising inflation. In periods following a fall in inflation, the impact of NFCI on the variance of inflation is far more muted. Overall, we can see how conditioning on momentum enriches our inference.

Note that these CPQR coefficients all have non-crossing constraints imposed. These constraints ensure that the fitted quantiles do not cross, and from \citet{szendrei2023fused} we know that such constraints lead to some regularisation on the coefficients across quantiles. As such, the quantile profiles would likely be even more pronounced had we not included these constraints. However, due to limited sample sizes, doing so would lead to more `jagged' quantile profiles. For completeness, we estimated the CPQR without non-crossing constraints and report these coefficient profiles in the appendix. The key takeaway is that the inclusion of non-crossing constraints do not drive our main findings.

\begin{figure}[t]
    \centering
    \includegraphics[width=\textwidth]{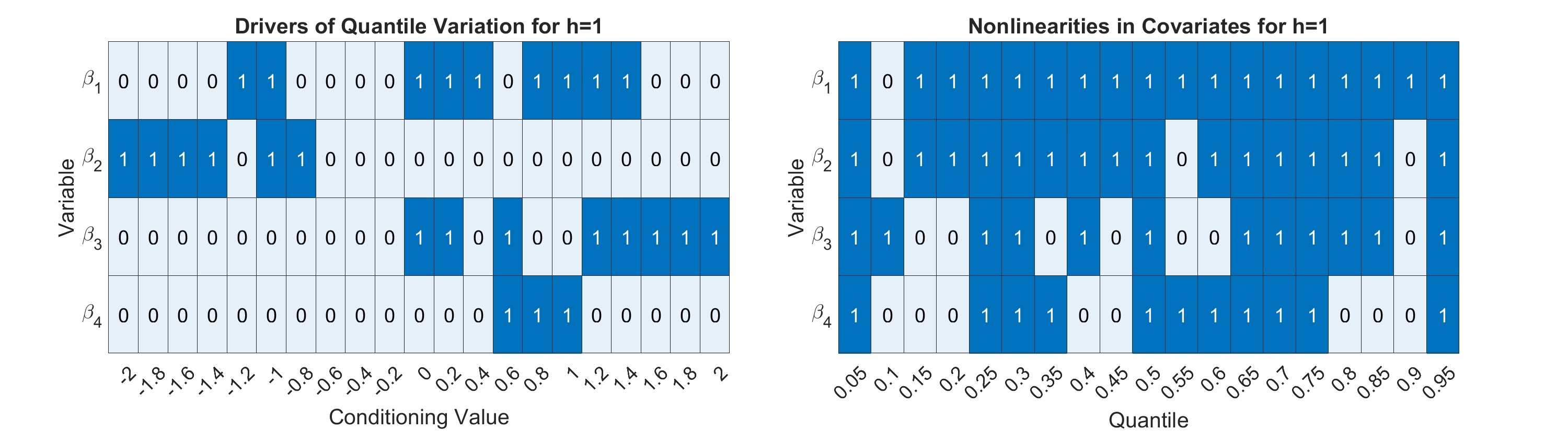}
    \caption{Drivers of nonlinearities and quantile variation for $h=1$}
    \label{fig:Hausman_h1}
\end{figure}

\begin{figure}[!h]
    \centering
    \includegraphics[width=\textwidth]{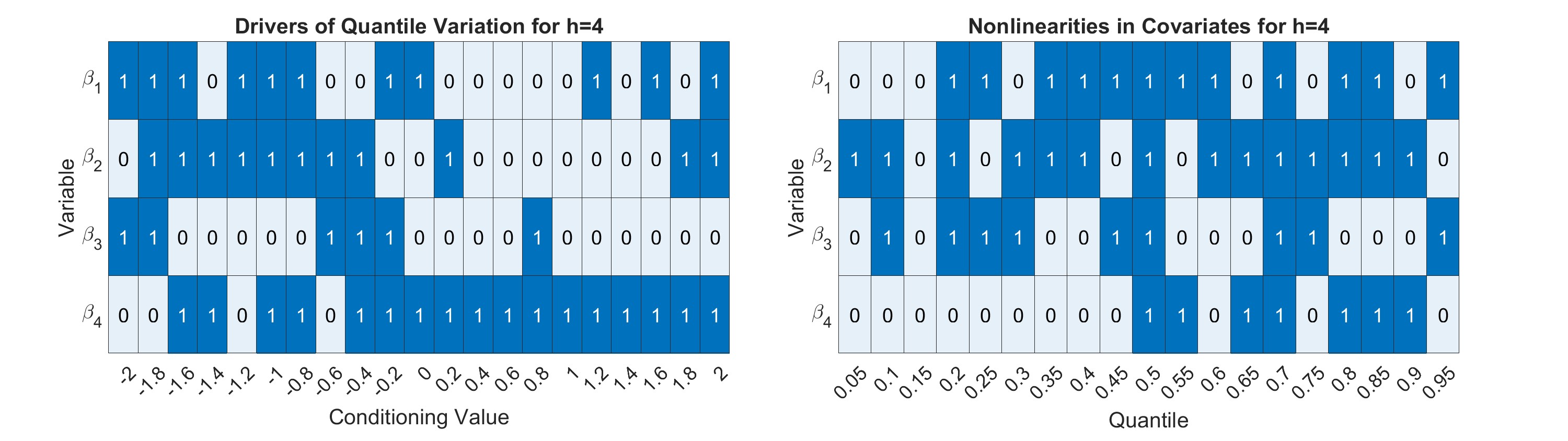}
    \caption{Drivers of nonlinearities and quantile variation for $h=4$}
    \label{fig:Hausman_h4}
\end{figure}

\subsection{Drivers of nonlinearities in inflation}

While the above coefficient profiles indicate quantile variation as well as conditioning on momentum of inflation, it does not inform precisely whether the variation induced by the specific nonlinearity is significant. To draw inference on what variables drive the different types of nonlinearities we conduct Hausman tests of the CPQR versus the CQR and QR. To visualise these findings we present the results of the Hausman tests in figures (\ref{fig:Hausman_h1}) and (\ref{fig:Hausman_h4}) for $h=1$ and $h=4$ respectively. In these figures a value of 1 represents where the CPQR is preferred. In the left panel the value 0 indicates that the CQR is preferred to CPQR, while in the right panel a value of 0 refers to the QR being preferred to CPQR; i.e. in either case, the zeroes indicate rejection of the corresponding restrictions by the Hausman test. Thus, the left panel refers to significant quantile variation while the right panel corresponds to significant momentum conditioning. These tests are conducted for each of the four covariates in turn, thus highlighting which variable drives nonlinearities or reflects quantile variation.


The figures are quite revealing about what type of nonlinearities are present for the different variables. For the shorter forecast horizon, in almost all cases the CPQR is preferred over the QR as seen in the right panel of figure (\ref{fig:Hausman_h1}). Interestingly, the same is not true for the CPQR when compared to the CQR for $h=1$ as seen in the left panel, which reflects that most of the nonlinearities are driven by conditioning on the momentum of inflation. 

There are some cases where both types of nonlinearities are important. The first is the impact of lagged inflation, when the momentum of inflation is near 0. Referring back to the contour plot of $\beta_1$ for $h=1$ in figure (\ref{fig:CPQRCoeff_h1}), shows that the contour plot reflects complex interaction along both dimensions. With the aid of Hausman test results, we can now infer that the increasing persistence in inflation at the upper quantiles is mostly true for increasing inflation periods. Another case where both nonlinearities are prominent is the GDP coefficient during falling inflation periods, which corroborates evidence from the contour plots, namely that the decreasing quantile profile of GDP growth is mostly present after a fall in inflation. Finally, there is evidence of a quantile profile for relative import prices during periods of rising  inflation. 

Taken together, the results highlight the trust in central banks target is a key characteristic in short run inflation dynamics, as evidenced by the right panel. The left panel, highlights that although the degree of trust in targets is important there still is a large degree of heterogeneity in how agents' form inflation expectations. In particular, when inflation changes are moderate, recent inflation values are more likely to be important for future inflation, which is in line with the simple forecasting strategy outlined by \citet{de2011animal}. However, as the change in inflation becomes more extreme, some agents' are less likely to look at past inflation. On account of this, pass through inflation (and the real sector through wage-setting) become an important source of nonlinearity as the degree to which agents' factor these alternative sources in their expectation formation is not uniform.


Figure (\ref{fig:Hausman_h4}) shows that the QR fares better than in the one quarter ahead forecasts. As such, quantile variation seems to be more important in capturing the nonlinear inflation dynamics. This is not surprising given that longer forecast horizons entail more uncertainty. As such, agents' inflation expectations are likely to be more diverse, which is captured by quantile regression. This is further evidenced by there being a larger overlap for CPQR being preferred in both dimensions of nonlinearity, indicating that for longer forecast horizons both types of nonlinearities are important. 

For lagged inflation CPQR provides better estimates in the extreme momentum cases (both sharply falling and rising inflation momentum) which is in stark contrast with the findings for $h=1$. This is likely on account of agents' believing that after extreme inflation movement, inflation will need longer time to return to the central banks target. The results for GDP growth in $h=1$ are shared with the longer forecast horizon: the quantile profile in GDP growth is mostly present in periods with falling inflation. For $\beta_3$, we see that CPQR is preferred over both QR and CQR mostly for near 0 stable inflation periods suggesting global factors play a more significant role in shaping longer run expectations when inflation is stable. Finally, the results for NFCI corroborate the findings of the contour plots in figure (\ref{fig:CPQRCoeff_h4}), namely that financial conditions evidence quantile variation for increasing inflation periods rather than quarters when inflation is falling.

Taking all the evidence in overall context, including the Hausman test and the coefficient profiles, it is interesting to note that the real sector (as captured by GDP growth) and the financial sector (as measured by NFCI) impact inflation very differently. In particular, there is quantile variation driven by the real sector in periods when inflation is falling, while in increasing inflation periods the financial sector has more influence on the distribution of future inflation but only in the longer horizon. Global factors (as captured by relative import prices) seem to be less important in driving the scale and skewness of inflation. Instead, it seems to act more as a location shifter, except for extreme inflation momentum values, where quantile variation is present.

The Hausman test results along with the coefficient profiles reinforce the importance of accounting for the momentum of inflation as it can be used as a way to gauge (lack of) trust in central bank targets during periods of significant changes in inflation. Furthermore, the fact that quantile variation becomes crucial for various covariates during periods of large changes in inflation for both horizons highlights how agents' inflation expectation formation becomes more heterogenous when they lose trust in the central banks ability to reach inflation targets. This highlights the need for policymakers to address the erosion of trust in central bank targets after extreme changes in inflation by taking proactive measures to restore credibility. This aligns with the conclusions of \citet{reis2023can}, who emphasise that reducing inflation after extreme spikes in inflation requires policy rates to exceed their neutral level.
\newpage

\subsection{In sample fit}

\begin{figure}[t]
    \centering
    \includegraphics[width=\textwidth]{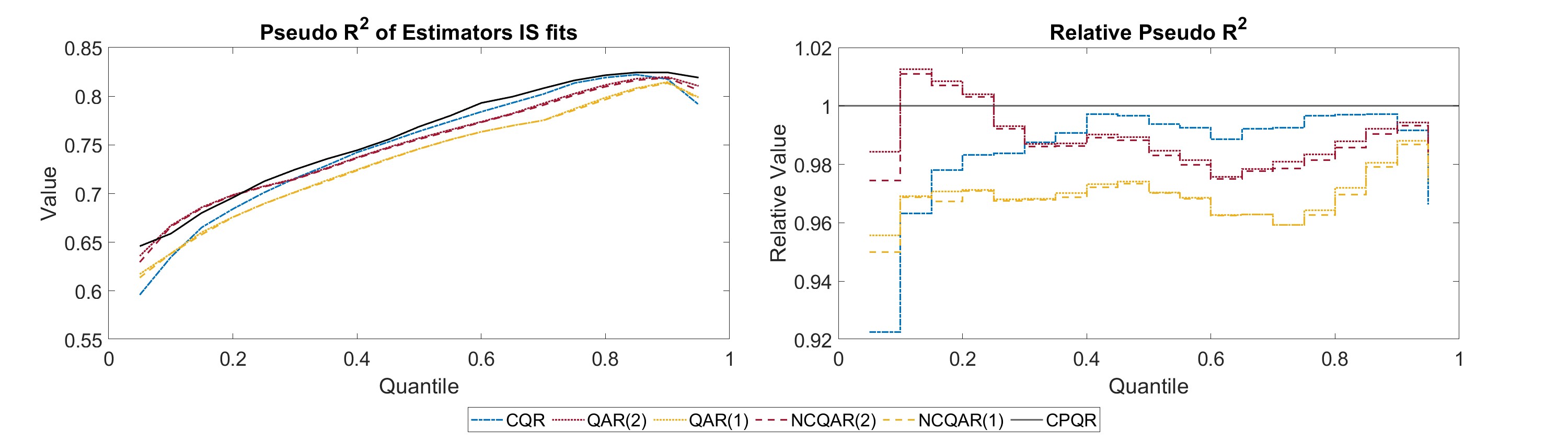}
    \caption{Pseudo $R^2$ of different estimators at different quantiles for $h=1$}
    \label{fig:R2_h1}
\end{figure}

\begin{figure}
    \centering
    \includegraphics[width=\textwidth]{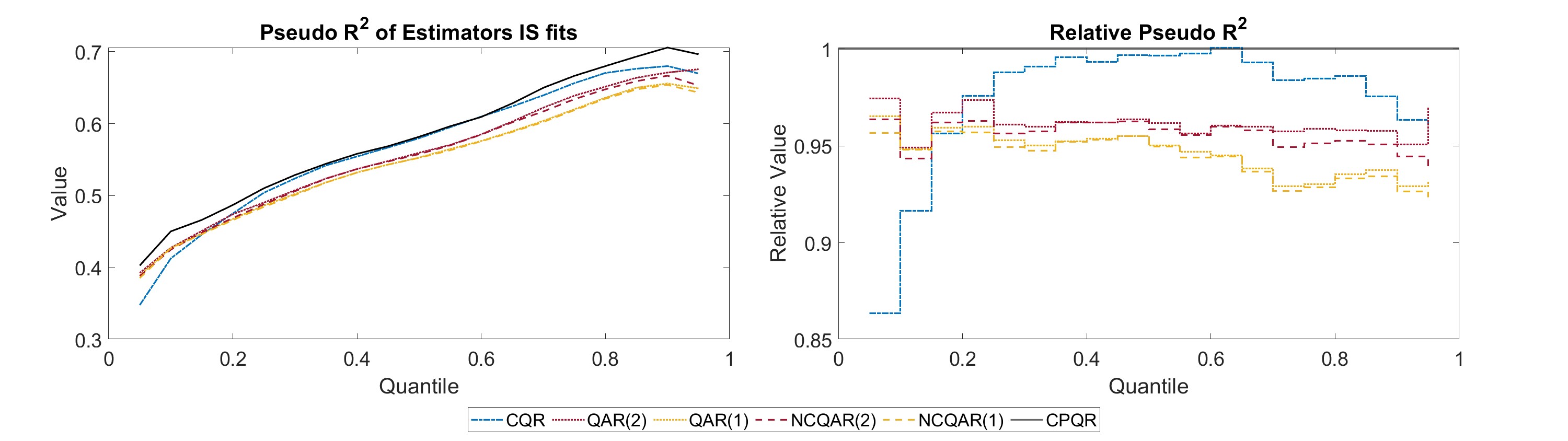}
    \caption{Pseudo $R^2$ of different estimators at different quantiles for $h=4$}
    \label{fig:R2_h4}
\end{figure}

To measure in sample fit we use the pseudo $R^2$ measure proposed by \citet{koenker1999goodness}. However, due to the conditionally parametric nature of the CPQR, calculating this measure directly is not possible. As such we use the method proposed in \citet{kohns23hsbqr}, i.e. to calculate the fitted quantiles first and then use these as covariates in a separate quantile regression. This idea is related to the VaR test of \citet{gaglianone2011evaluating}. In essence, the better the quantile fits the data, the more likely it will provide all the information necessary for a quantile regression with only the fitted quantile as a covariate (and an intercept). Thus, we calculate the pseudo $R^2$ using the following regression:
\begin{equation}
    \label{eq:backtest5}
    Q_{y_{t+h}}(\tau|V_{t+h,\tau})=\theta_0+\theta_1V_{t+h,\tau},
\end{equation}
where $V_{t+h,\tau}$ is the fitted value (prediction) for the $\tau^{th}$ quantile. First we calculate the fitted values for all estimators, and then use this as the only covariate in equation (\ref{eq:backtest5}). Then the pseudo $R^2$ is calculated by:
\begin{equation}
    R^2=1-\frac{RASW}{TASW},
\end{equation}
where $RASW$ is the residual absolute sum of weighted differences (or, residuals) of equation (\ref{eq:backtest5}). $TASW$ is the total absolute sum of weighted differences, which constitute the residuals of equation (\ref{eq:backtest5}) where $\theta_1$ is constrained to 0. In essence, the pseudo $R^2$ measures how much information the fitted value adds to the regression compared to a quantile regression with just the intercept. The results of this exercise are reported in figure (\ref{fig:R2_h1}) and (\ref{fig:R2_h4}), for the one-quarter ahead and one-year ahead fits respectively. The left panel shows the raw $R^2$ of several competing estimators, while the right panel shows the estimators' $R^2$ relative to the CPQR.

The pseudo $R^2$ of competing estimators show that $V_{t+h,\tau}$ brings greater improvement at the upper quantiles of inflation. This is true for all estimators, which highlights that the lower quantiles of inflation can be particularly hard to model. Considering the relative performance of the different estimators, we find that the CPQR has the best performance at the tails and the median for both forecast horizons. Importantly, the CPQR beats both QAR(2) and CQR for most quantiles. The QAR(2) model provides better fit than the CPQR only for the intermediate lower quantiles of inflation when $h=1$. However, this is not the case for the longer forecast horizon, where the CPQR yields better fit than the QAR(2) for every quantile. Interestingly, the CQR also beats the QAR(2) model at the central quantiles, highlighting that conditioning on the momentum of inflation is an important source of nonlinearity that drives inflation.

For completeness we also estimate the QAR(1) model, but as expected, it always yields worse performance than the QAR(2). Furthermore, we estimate non-crossing constraint variants of these models (denoted as NCQAR(1) and NCQAR(2)), which only offer marginally worse fits than their counterparts without such constraints. Since these figures only focus on in sample fit, it is not surprising that the non-crossing constraints do not improve fit; see also \citet{szendrei2023fused}.

\subsection{Out of sample fit}

\begin{figure}
    \centering
    \includegraphics[width=\textwidth]{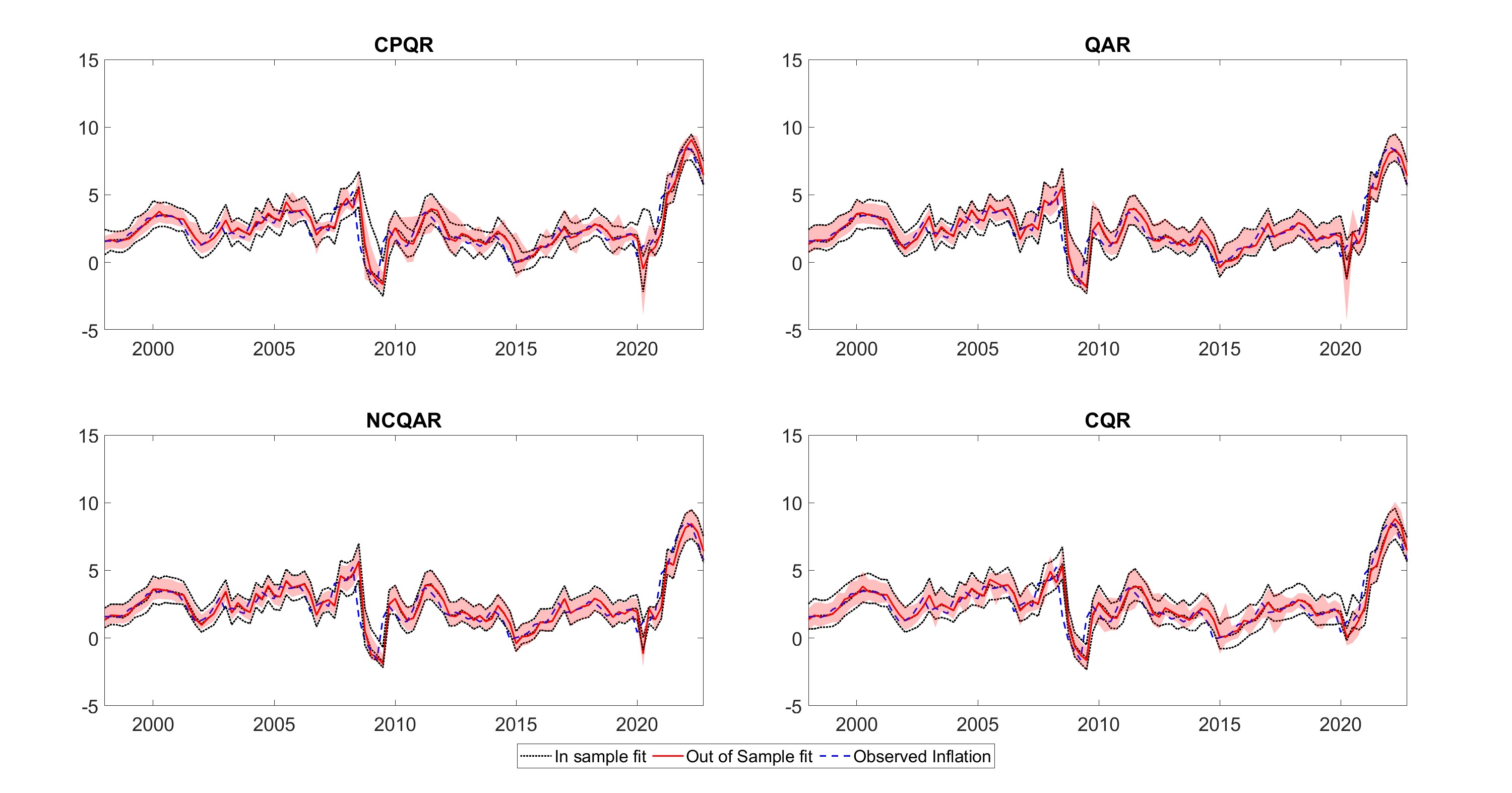}
    \caption{Out of Sample fits of different estimators for $h=1$}
    \label{fig:OOS_h1}
\end{figure}

\begin{figure}
    \centering
    \includegraphics[width=\textwidth]{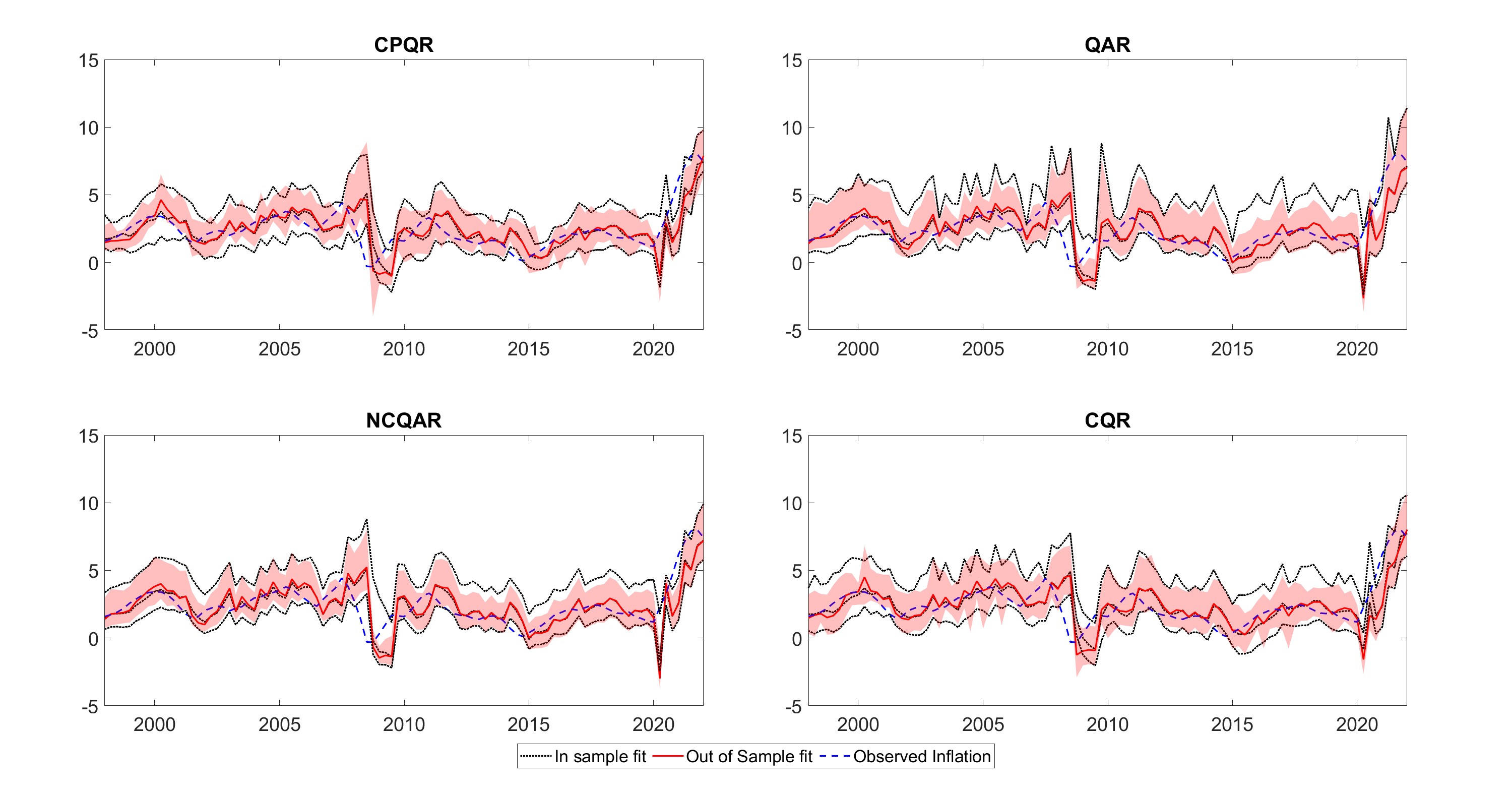}
    \caption{Out of Sample fits of different estimators for $h=4$}
    \label{fig:OOS_h4}
\end{figure}

Finally, we compare the out-of-sample forecast performance of the CPQR compared to the the QAR(2), NCQAR(2), and CQR. For this exercise we first estimate using 100 observations with the window expanding by 1 period each time. The estimated $\beta$'s are then used to obtain a one-quarter ahead (and one year ahead) forecasts. The initial period is set longer than \citet{adrian2019vulnerable} because CPQR estimation needs more data. Figures (\ref{fig:OOS_h1}) and (\ref{fig:OOS_h4}) show the observed inflation, and the out-of-sample fits compared to the in-sample fits for three quantiles ($5^{th}$, $50^{th}$, and $95^{th}$), at $h=1$ and $h=4$, respectively. When evaluating out-of-sample fits, we sort the forecasted quantiles of all estimators using the procedure of \citet{chernozhukov2010quantile}. It is common in macroeconomic at-risk model forecasting to either apply a sorting procedure, or to fit a distribution using the fitted quantiles as inputs. This helps the traditional QR estimator the most, but it also improves the fit for the other estimators with non-crossing constraints. As discussed in \citet{szendrei2023fused}, this is because these constraints only impose monotonically increasing quantiles in-sample, but not out-of-sample.

The different fits for $h=1$ in figure (\ref{fig:OOS_h1}) reveal that there is not much to distinguish between the different estimators. However, three key observations emerge. First, the QAR and NCQAR out of sample fits over time are not much different from their in-sample fits over the whole period. Second, the CPQR and CQR forecast wider inflation density during the 2008 period compared to the NCQAR. Third, during the fall in inflation in 2019, estimators that allow for quantile variation forecast the potential for deflation, while the CQR's lowest forecast was of no inflation.

The longer forecast horizon (figure \ref{fig:OOS_h4}) reveals larger differences. The QAR's forecast and fitted values deviate considerably, especially at the upper quantiles. Although less pronounced, the same is true for the NCQAR. This shows how imposing non-crossing constraints can help in forecasting, but also not necessarily enough if the model is misspecified. The CQR also shows this problem for 2015 and onwards. The CPQR on the other hand provides out-of-sample fits that are closer to the in-sample fits for the aforementioned period. This highlights the need to model both types of nonlinearities (conditioning on momentum of inflation and quantile variation) in IaR. A second period where the CPQR provides better density estimates is the 2008 global financial crisis. The out of sample density estimates for all other estimators show a sudden downward shift in the density of inflation, but simultaneously also a less spread density. The CPQR on the other hand forecast a lower inflation location without shrinking the forecast inflation density.

\begin{table}[]
\centering
\resizebox{\columnwidth}{!}{%
\begin{tabular}{l|cccc|cccc}
\hline
 & \multicolumn{4}{c|}{$h=1$} & \multicolumn{4}{c}{$h=4$} \\
 & CPQR & QAR(2) & NCQAR(2) & CQR & CPQR & QAR(2) & NCQAR(2) & CQR \\ \hline
$w_q^1$ & 0.221 & 0.222 & 0.222 & 0.221 & 0.326* & 0.363 & 0.362 & 0.330** \\
$w_q^2$ & 0.042 & 0.043 & 0.043 & 0.042 & 0.063* & 0.069 & 0.069 & 0.064** \\
$w_q^3$ & 0.067 & 0.066 & 0.067 & 0.067 & 0.097* & 0.105 & 0.104 & 0.098** \\
$w_q^4$ & 0.069 & 0.070 & 0.070 & 0.070 & 0.104 & 0.119 & 0.119 & 0.105** \\
\hline
\end{tabular}
}%
\caption{qwCRPS for the different weight profiles. Stars represent significance at the 10\% (*), 5\% (**), and 1\% (***) level respectively.}
\label{tab:qwCRPS}
\end{table}

Beyond visually inspecting the forecast densities (and the in-sample fitted densities), we also evaluate forecast performance of the different estimators using the quantile weighted CRPS (qwCRPS) of \citet{gneiting2011comparing} as a scoring rule. To calculate this measure, we first take the Quantile Score (QS), which is the weighted residual for a given forecast observation, $\hat{y}_{t+h,p}$. The qwCRPS is then calculated as:

\begin{equation}
    qwCRPS_{t+h} = \int^1_0 \; w_q QS_{t+h,q}dq,
\end{equation}
where $w_q$ denotes a weighting scheme to evaluate specific parts of the forecast density. This scoring rule allows evaluating different part of the distribution by using different weighting schemes. We consider 4 different weighting schemes: $w_q^1=\frac{1}{Q}$ places equal weight on all quantiles;\footnote{This is equivalent to taking the average of the weighted residuals at a given observation.} $w_q^2=q(1-q)$ places more weight on central quantiles; $w_q^3=(1-q)^2$ places more weight on the left tail; and $w_q^4=q^2$ more weight on the right tail. The results using the different weighting schemes are presented in Table (\ref{tab:qwCRPS}). To check whether the differences are statistically significant, the Diebold-Mariano test is used with the QAR(2) estimator as a reference \citep{diebold1995comparing}. 

Table (\ref{tab:qwCRPS}) reveals that the CPQR provides the best forecast performance among the estimators for almost all weights. The only case where this is not true is the left tail for the shorter forecast horizon. Nevertheless, the improvements in forecast performance are marginal at the short forecast horizon, with the Diebold-Mariano test showing no significant difference between the CPQR and the QAR(2). This is not true for the longer forecast horizon, where the CPQR yields significantly better forecast performance than the QAR(2) for all weights considered except at the right tail. However, statistical significance is achieved only at the 10\% level because of the considerably larger variation modelled by CPQR and correspondingly larger data that the estimator requires. Importantly, the CQR also yields improved forecast results, statistically significant at the 5\% level, even if the forecasts are marginally worse than the CPQR. Overall, this highlights the importance of conditioning on the momentum of inflation for longer forecast horizons of inflation density.

\begin{figure}
    \centering
    \includegraphics[width=\textwidth]{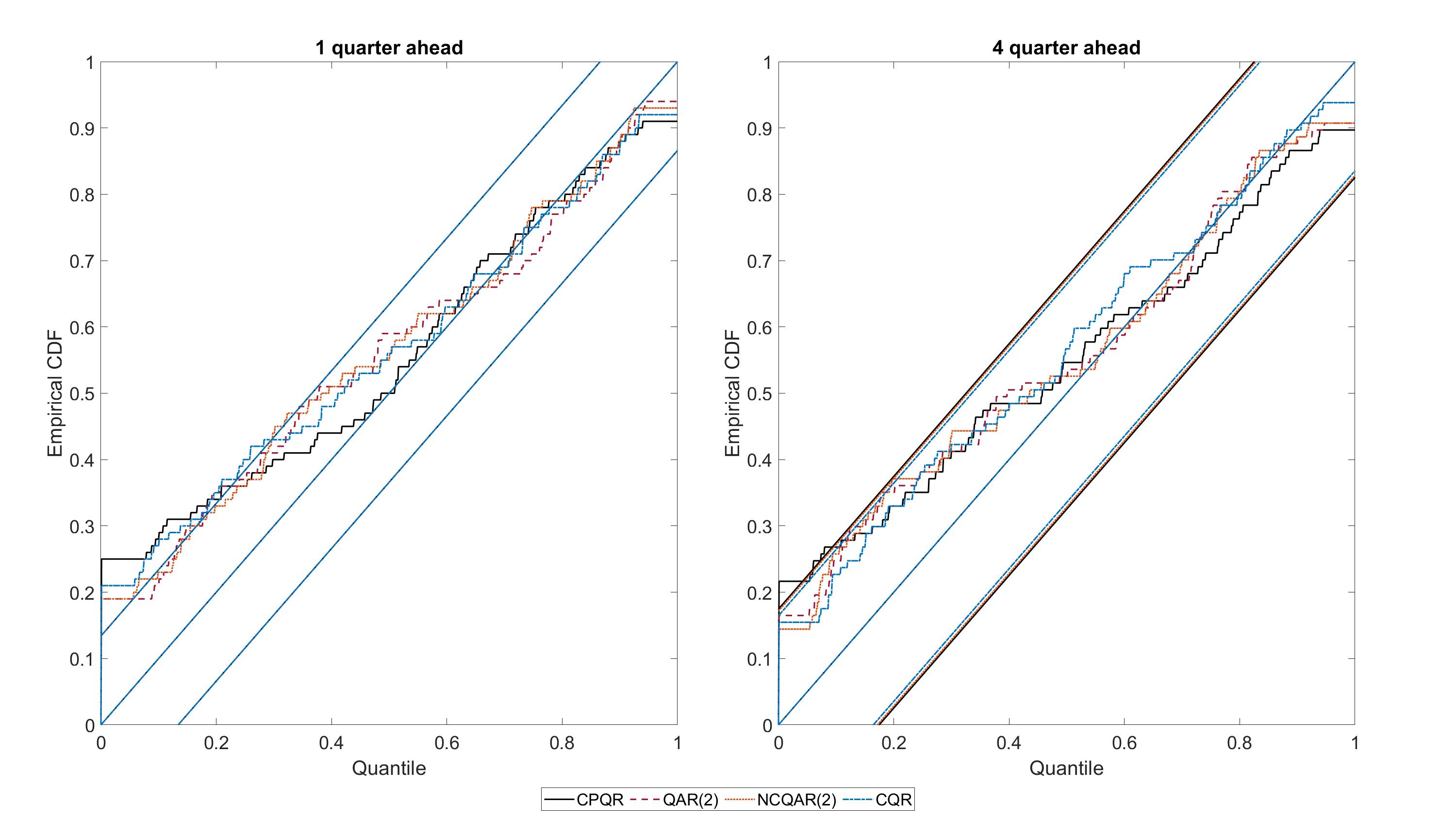}
    \caption{PITS of different estimators}
    \label{fig:PITS}
\end{figure}

Since the qwCRPS does not inform about forecast calibration across different parts of the overall forecast distribution, we also apply the \citet{rossi2019alternative} calibration test. The idea behind the test is that if a model is well calibrated, then CDF of the probability integral transforms (PITs) should fall along the 45 degree line \citep{diebold1998vevaluating}. We also report confidence bands around this 45 degree baseline to account for sampling uncertainty \citep{rossi2019alternative}. Note that these confidence bands are provided for general guidance only, since they are derived using rolling window estimates rather than an expanding window. The PITs of out-of-sample forecasts for the different estimators are reported in figure (\ref{fig:PITS}): left panel -- one-quarter ahead and right panel -- one year ahead forecasts.

For the one quarter ahead horizon (left panel) we find that all estimators breach the boundary at the lower quantiles. This suggests that the extreme left tails of the distribution are not well-calibrated but the rest of the distribution performs better. This is in line with what we find in the in-sample results: the left tail of the fitted inflation density has lower pseudo $R^2$ than the right tail. However, the trajectory of the CPQR across quantiles shows a convergence towards the 45 degree line earlier than the other estimators. Furthermore, the PITs of the CPQR align with the 45 degree line almost perfectly from the median onwards. This suggests that just like the in-sample results, the centre and right tail of the out-of-sample forecasted distribution are better captured by the model.

For the one year ahead horizon (right panel) there is less difference between the estimators and are in general better calibrated. The only exception is the CPQR at the very extreme low quantile, where the PIT breaches the boundary. Given that these boundaries are reported merely for guidance, and that the CPQR tends towards the 45 degree line, we conclude that the CPQR provides well calibrated densities the longer forecast horizons as well. Just like in the case of the short horizon (and in-sample), the estimators seem to fare better for the central and right quantiles. As such, one avenue of future research is to find models that can provide better calibrated density fits even at the left tail of the inflation distribution.


In summary, the out of sample performance of the CPQR corroborates the findings of coefficient profiles, that quantile variation is not the only important nonlinearity when it comes to estimating inflation densities. In particular, conditioning on the momentum of inflation provides valuable information which helps in obtaining better forecast performance, especially at longer forecast horizons.

\section{Conclusion}

This paper extended \citet{adrian2019vulnerable} in estimating Inflation-at-Risk. Compared to other IaR estimates such as \citet{banerjee2020inflation} and \citet{lopez2022inflation}, we introduce an additional source of nonlinearity, namely conditioning on the momentum of inflation. We do so because economic agents' expectations are likely to be influenced by the momentum of inflation, which in turn leads to different monetary policy reactions. Since combining conditioning and quantile regression restricts the effective sample size, we formulate non-crossing constraints imposed locally for the conditioning value. This novel approach creates coefficient profiles across quantiles as well as across the different values of inflation momentum.

Using our proposed CPQR (conditionally parametric quantile regression) method with non-crossing constraints, we estimate a IaR model with 4 covariates: lagged inflation, lagged GDP growth, lagged relative import price inflation, and lagged NFCI. Using this IaR model we show that there is evidence for both types of nonlinearities when applied to US data between 1973 and 2023. Importantly, we find that the real sector (as captured by GDP) and the financial sector (as measured by NFCI) impact inflation very differently. In particular, there is quantile variation which is driven by the real sector in periods of falling inflation, while in increasing inflation periods the financial sector has more influence on the distribution of future inflation. Global factors (as captured by relative import prices) are less important in driving the shape of inflation. Instead, they have a larger impact as a location shifter, depending on the value of the momentum of inflation.

Evaluating the performance of the CPQR also provides evidence of the two types of nonlinearities driving IaR. In particular, we find that pseudo $R^2$ of the CPQR is the best among alternate estimators considered, particularly at the tails of the conditional distribution. We also find that the CQR (composite quantile regression) provides good results at the central quantiles, showing that the location of the inflation distribution is likely more driven by conditioning on the momentum of inflation. The out-of-sample performance of the estimators corroborates the findings of the in-sample fit exercises. We find that at the longer forecast horizon, the CPQR and CQR provide significantly better forecast performance than the QAR(2) (quantile autoregressive model with 2 lags).

In this paper we only focused on implementing constraints across quantiles but not across the conditioning variable grid. Due to this the quantile profiles can be smoothed, but moving from one grid point to the next can have large shifts in the estimated coefficients. An avenue for future research is to implement cross grid constraints as well to achieve a smooth profile across both quantiles and conditioning variable grid.

In the paper we also found that the models considered fit the central and upper quantiles of inflation much better than the lower quantiles. This discrepancy may stem from the variables selected to describe the inflation density. Future research could focus on improving the modelling of the left tail of inflation density to address this.

\pagebreak

\bibliographystyle{chicago}
\bibliography{main.bbl}

\begin{thebibliography}{}

\bibitem[\protect\citeauthoryear{Adrian, Boyarchenko, and Giannone}{Adrian et~al.}{2019}]{adrian2019vulnerable}
Adrian, T., N.~Boyarchenko, and D.~Giannone (2019).
\newblock Vulnerable growth.
\newblock {\em American Economic Review\/}~{\em 109\/}(4), 1263--89.

\bibitem[\protect\citeauthoryear{Adrian, Boyarchenko, and Giannone}{Adrian et~al.}{2021}]{adrian2021multimodality}
Adrian, T., N.~Boyarchenko, and D.~Giannone (2021).
\newblock Multimodality in macrofinancial dynamics.
\newblock {\em International Economic Review\/}~{\em 62\/}(2), 861--886.

\bibitem[\protect\citeauthoryear{Akerlof and Shiller}{Akerlof and Shiller}{2010}]{akerlof2010animal}
Akerlof, G.~A. and R.~J. Shiller (2010).
\newblock {\em Animal spirits: How human psychology drives the economy, and why it matters for global capitalism}.
\newblock Princeton university press.

\bibitem[\protect\citeauthoryear{Banerjee, Contreras, Mehrotra, and Zampolli}{Banerjee et~al.}{2020}]{banerjee2020inflation}
Banerjee, R.~N., J.~Contreras, A.~Mehrotra, and F.~Zampolli (2020).
\newblock Inflation at risk in advanced and emerging market economies.

\bibitem[\protect\citeauthoryear{Bates, Hastie, and Tibshirani}{Bates et~al.}{2024}]{bates2024cross}
Bates, S., T.~Hastie, and R.~Tibshirani (2024).
\newblock Cross-validation: what does it estimate and how well does it do it?
\newblock {\em Journal of the American Statistical Association\/}~{\em 119\/}(546), 1434--1445.

\bibitem[\protect\citeauthoryear{Blanchard, Cerutti, and Summers}{Blanchard et~al.}{2015}]{blanchard2015inflation}
Blanchard, O., E.~Cerutti, and L.~Summers (2015).
\newblock Inflation and activity--two explorations and their monetary policy implications.
\newblock Technical report, National Bureau of Economic Research.

\bibitem[\protect\citeauthoryear{Bondell, Reich, and Wang}{Bondell et~al.}{2010}]{bondell2010noncrossing}
Bondell, H.~D., B.~J. Reich, and H.~Wang (2010).
\newblock Noncrossing quantile regression curve estimation.
\newblock {\em Biometrika\/}~{\em 97\/}(4), 825--838.

\bibitem[\protect\citeauthoryear{Brave and Butters}{Brave and Butters}{2018}]{brave2018diagnosing}
Brave, S. and R.~A. Butters (2018).
\newblock Diagnosing the financial system: financial conditions and financial stress.
\newblock {\em 29th issue (June 2012) of the International Journal of Central Banking\/}.

\bibitem[\protect\citeauthoryear{Chaudhuri}{Chaudhuri}{1991}]{chaudhuri1991nonparametric}
Chaudhuri, P. (1991).
\newblock Nonparametric estimates of regression quantiles and their local bahadur representation.
\newblock {\em The Annals of statistics\/}~{\em 19\/}(2), 760--777.

\bibitem[\protect\citeauthoryear{Chen, Dolado, and Gonzalo}{Chen et~al.}{2021}]{chen2021quantile}
Chen, L., J.~J. Dolado, and J.~Gonzalo (2021).
\newblock Quantile factor models.
\newblock {\em Econometrica\/}~{\em 89\/}(2), 875--910.

\bibitem[\protect\citeauthoryear{Chernozhukov, Fern{\'a}ndez-Val, and Galichon}{Chernozhukov et~al.}{2010}]{chernozhukov2010quantile}
Chernozhukov, V., I.~Fern{\'a}ndez-Val, and A.~Galichon (2010).
\newblock Quantile and probability curves without crossing.
\newblock {\em Econometrica\/}~{\em 78\/}(3), 1093--1125.

\bibitem[\protect\citeauthoryear{Cho and Rho}{Cho and Rho}{2023}]{choreassessing}
Cho, D. and S.~Rho (2023).
\newblock Reassessing growth vulnerability.
\newblock {\em Journal of Applied Econometrics\/}~{\em n/a\/}(n/a).

\bibitem[\protect\citeauthoryear{Clark, Huber, Koop, Marcellino, and Pfarrhofer}{Clark et~al.}{2023}]{clark2023tail}
Clark, T.~E., F.~Huber, G.~Koop, M.~Marcellino, and M.~Pfarrhofer (2023).
\newblock Tail forecasting with multivariate bayesian additive regression trees.
\newblock {\em International Economic Review\/}~{\em 64\/}(3), 979--1022.

\bibitem[\protect\citeauthoryear{Cleveland}{Cleveland}{1994}]{cleveland1994coplots}
Cleveland, W.~S. (1994).
\newblock Coplots, nonparametric regression, and conditionally parametric fits.
\newblock {\em Lecture Notes-Monograph Series\/}, 21--36.

\bibitem[\protect\citeauthoryear{De~Grauwe}{De~Grauwe}{2011}]{de2011animal}
De~Grauwe, P. (2011).
\newblock Animal spirits and monetary policy.
\newblock {\em Economic theory\/}~{\em 47}, 423--457.

\bibitem[\protect\citeauthoryear{Diebold, Gunther, and Tay}{Diebold et~al.}{1998}]{diebold1998vevaluating}
Diebold, F.~X., T.~A. Gunther, and A.~S. Tay (1998).
\newblock Evaluating density forecasts with applications to financial risk management.
\newblock {\em International Economic Review\/}~{\em 39\/}(4), 863--883.

\bibitem[\protect\citeauthoryear{Diebold and Mariano}{Diebold and Mariano}{1995}]{diebold1995comparing}
Diebold, F.~X. and R.~S. Mariano (1995).
\newblock Comparing predictive accuracy.
\newblock {\em Journal of Business \& Economic Statistics\/}~{\em 13\/}(3).

\bibitem[\protect\citeauthoryear{Enders and Granger}{Enders and Granger}{1998}]{enders1998unit}
Enders, W. and C.~W.~J. Granger (1998).
\newblock Unit-root tests and asymmetric adjustment with an example using the term structure of interest rates.
\newblock {\em Journal of Business \& Economic Statistics\/}~{\em 16\/}(3), 304--311.

\bibitem[\protect\citeauthoryear{Enders and Siklos}{Enders and Siklos}{2001}]{enders2001cointegration}
Enders, W. and P.~L. Siklos (2001).
\newblock Cointegration and threshold adjustment.
\newblock {\em Journal of Business \& Economic Statistics\/}~{\em 19\/}(2), 166--176.

\bibitem[\protect\citeauthoryear{Ferrara, Mogliani, and Sahuc}{Ferrara et~al.}{2022}]{ferrara2022high}
Ferrara, L., M.~Mogliani, and J.-G. Sahuc (2022).
\newblock High-frequency monitoring of growth at risk.
\newblock {\em International Journal of Forecasting\/}~{\em 38\/}(2), 582--595.

\bibitem[\protect\citeauthoryear{Figueres and Jaroci{\'n}ski}{Figueres and Jaroci{\'n}ski}{2020}]{figueres2020vulnerable}
Figueres, J.~M. and M.~Jaroci{\'n}ski (2020).
\newblock Vulnerable growth in the euro area: Measuring the financial conditions.
\newblock {\em Economics Letters\/}~{\em 191}, 109126.

\bibitem[\protect\citeauthoryear{Gaglianone, Lima, Linton, and Smith}{Gaglianone et~al.}{2011}]{gaglianone2011evaluating}
Gaglianone, W.~P., L.~R. Lima, O.~Linton, and D.~R. Smith (2011).
\newblock Evaluating value-at-risk models via quantile regression.
\newblock {\em Journal of Business \& Economic Statistics\/}~{\em 29\/}(1), 150--160.

\bibitem[\protect\citeauthoryear{Gal{\'a}n}{Gal{\'a}n}{2020}]{galan2020benefits}
Gal{\'a}n, J.~E. (2020).
\newblock The benefits are at the tail: uncovering the impact of macroprudential policy on growth-at-risk.
\newblock {\em Journal of Financial Stability\/}, 100831.

\bibitem[\protect\citeauthoryear{Galvao, Montes-Rojas, and Olmo}{Galvao et~al.}{2011}]{galvao2011threshold}
Galvao, A.~F., G.~Montes-Rojas, and J.~Olmo (2011).
\newblock Threshold quantile autoregressive models.
\newblock {\em Journal of Time Series Analysis\/}~{\em 32\/}(3), 253--267.

\bibitem[\protect\citeauthoryear{Gneiting and Ranjan}{Gneiting and Ranjan}{2011}]{gneiting2011comparing}
Gneiting, T. and R.~Ranjan (2011).
\newblock Comparing density forecasts using threshold-and quantile-weighted scoring rules.
\newblock {\em Journal of Business \& Economic Statistics\/}~{\em 29\/}(3), 411--422.

\bibitem[\protect\citeauthoryear{Hausman}{Hausman}{1978}]{hausman1978specification}
Hausman, J.~A. (1978).
\newblock Specification tests in econometrics.
\newblock {\em Econometrica: Journal of the econometric society\/}, 1251--1271.

\bibitem[\protect\citeauthoryear{Jord{\`a} and Nechio}{Jord{\`a} and Nechio}{2023}]{jorda2023inflation}
Jord{\`a}, {\`O}. and F.~Nechio (2023).
\newblock Inflation and wage growth since the pandemic.
\newblock {\em European Economic Review\/}~{\em 156}, 104474.

\bibitem[\protect\citeauthoryear{Kahneman and Thaler}{Kahneman and Thaler}{2006}]{kahneman2006anomalies}
Kahneman, D. and R.~H. Thaler (2006).
\newblock Anomalies: Utility maximization and experienced utility.
\newblock {\em Journal of economic perspectives\/}~{\em 20\/}(1), 221--234.

\bibitem[\protect\citeauthoryear{Koenker}{Koenker}{1984}]{koenker1984note}
Koenker, R. (1984).
\newblock A note on l-estimates for linear models.
\newblock {\em Statistics \& probability letters\/}~{\em 2\/}(6), 323--325.

\bibitem[\protect\citeauthoryear{Koenker}{Koenker}{2005}]{koenker2005}
Koenker, R. (2005).
\newblock {\em Quantile regression}.
\newblock New York: Cambridge University Press.

\bibitem[\protect\citeauthoryear{Koenker and Bassett}{Koenker and Bassett}{1978}]{koenker1978regression}
Koenker, R. and G.~Bassett (1978).
\newblock Regression quantiles.
\newblock {\em Econometrica: journal of the Econometric Society\/}, 33--50.

\bibitem[\protect\citeauthoryear{Koenker and Machado}{Koenker and Machado}{1999}]{koenker1999goodness}
Koenker, R. and J.~A. Machado (1999).
\newblock Goodness of fit and related inference processes for quantile regression.
\newblock {\em Journal of the american statistical association\/}~{\em 94\/}(448), 1296--1310.

\bibitem[\protect\citeauthoryear{Koenker and Xiao}{Koenker and Xiao}{2006}]{koenker2006quantile}
Koenker, R. and Z.~Xiao (2006).
\newblock Quantile autoregression.
\newblock {\em Journal of the American statistical association\/}~{\em 101\/}(475), 980--990.

\bibitem[\protect\citeauthoryear{Kohns and Szendrei}{Kohns and Szendrei}{2021}]{kohns2021decoupling}
Kohns, D. and T.~Szendrei (2021).
\newblock Decoupling shrinkage and selection for the bayesian quantile regression.
\newblock {\em arXiv preprint arXiv:2107.08498\/}.

\bibitem[\protect\citeauthoryear{Kohns and Szendrei}{Kohns and Szendrei}{2023}]{kohns23hsbqr}
Kohns, D. and T.~Szendrei (2023).
\newblock {Horseshoe prior Bayesian quantile regression}.
\newblock {\em Journal of the Royal Statistical Society Series C: Applied Statistics\/}, qlad091.

\bibitem[\protect\citeauthoryear{Korobilis}{Korobilis}{2017}]{korobilis2017quantile}
Korobilis, D. (2017).
\newblock Quantile regression forecasts of inflation under model uncertainty.
\newblock {\em International Journal of Forecasting\/}~{\em 33\/}(1), 11--20.

\bibitem[\protect\citeauthoryear{Lenza, Moutachaker, and Paredes}{Lenza et~al.}{2023}]{lenza2023density}
Lenza, M., I.~Moutachaker, and J.~Paredes (2023).
\newblock Density forecasts of inflation: a quantile regression forest approach.

\bibitem[\protect\citeauthoryear{Lieb and Schuffels}{Lieb and Schuffels}{2022}]{lieb2022inflation}
Lieb, L. and J.~Schuffels (2022).
\newblock Inflation expectations and consumer spending: the role of household balance sheets.
\newblock {\em Empirical Economics\/}~{\em 63\/}(5), 2479--2512.

\bibitem[\protect\citeauthoryear{Lopez-Salido and Loria}{Lopez-Salido and Loria}{2022}]{lopez2022inflation}
Lopez-Salido, D. and F.~Loria (2022).
\newblock Inflation at risk.
\newblock {\em Available at SSRN 4002673\/}.

\bibitem[\protect\citeauthoryear{Loria, Cascaldi-Garcia, Cuba-Borda, and Caldara}{Loria et~al.}{2020}]{loria2020understanding}
Loria, F., D.~Cascaldi-Garcia, P.~Cuba-Borda, and D.~Caldara (2020).
\newblock Understanding growth-at-risk: A markov switching approach.
\newblock {\em Available at SSRN\/}.

\bibitem[\protect\citeauthoryear{McMillen}{McMillen}{1996}]{mcmillen1996one}
McMillen, D.~P. (1996).
\newblock One hundred fifty years of land values in chicago: A nonparametric approach.
\newblock {\em Journal of Urban Economics\/}~{\em 40\/}(1), 100--124.

\bibitem[\protect\citeauthoryear{McMillen}{McMillen}{2012}]{mcmillen2012quantile}
McMillen, D.~P. (2012).
\newblock {\em Quantile regression for spatial data}.
\newblock Springer Science \& Business Media.

\bibitem[\protect\citeauthoryear{McMillen}{McMillen}{2015}]{mcmillen2015conditionally}
McMillen, D.~P. (2015).
\newblock Conditionally parametric quantile regression for spatial data: An analysis of land values in early nineteenth century chicago.
\newblock {\em Regional Science and Urban Economics\/}~{\em 55}, 28--38.

\bibitem[\protect\citeauthoryear{Mitchell, Zhu, and Poon}{Mitchell et~al.}{2022}]{mitchell2022constructing}
Mitchell, J., D.~Zhu, and A.~Poon (2022).
\newblock Constructing density forecasts from quantile regressions: Multimodality in macro-financial dynamics.

\bibitem[\protect\citeauthoryear{Pfarrhofer}{Pfarrhofer}{2022}]{pfarrhofer2022modeling}
Pfarrhofer, M. (2022).
\newblock Modeling tail risks of inflation using unobserved component quantile regressions.
\newblock {\em Journal of Economic Dynamics and Control\/}~{\em 143}, 104493.

\bibitem[\protect\citeauthoryear{Racine and Li}{Racine and Li}{2004}]{racine2004nonparametric}
Racine, J. and Q.~Li (2004).
\newblock Nonparametric estimation of regression functions with both categorical and continuous data.
\newblock {\em Journal of Econometrics\/}~{\em 119\/}(1), 99--130.

\bibitem[\protect\citeauthoryear{Reis}{Reis}{2023}]{reis2023can}
Reis, R. (2023).
\newblock What can keep euro area inflation high?
\newblock {\em Economic Policy\/}~{\em 38\/}(115), 495--517.

\bibitem[\protect\citeauthoryear{Rossi and Sekhposyan}{Rossi and Sekhposyan}{2019}]{rossi2019alternative}
Rossi, B. and T.~Sekhposyan (2019).
\newblock Alternative tests for correct specification of conditional predictive densities.
\newblock {\em Journal of Econometrics\/}~{\em 208\/}(2), 638--657.

\bibitem[\protect\citeauthoryear{Suarez}{Suarez}{2022}]{suarez2022growth}
Suarez, J. (2022).
\newblock Growth-at-risk and macroprudential policy design.
\newblock {\em Journal of Financial Stability\/}~{\em 60}, 101008.

\bibitem[\protect\citeauthoryear{Szendrei, Bhattacharjee, and Schaffer}{Szendrei et~al.}{2024}]{szendrei2023fused}
Szendrei, T., A.~Bhattacharjee, and M.~E. Schaffer (2024).
\newblock Fused {LASSO} as non-crossing quantile regression.
\newblock {\em arXiv preprint arXiv:2403.14036\/}.

\bibitem[\protect\citeauthoryear{Szendrei and Varga}{Szendrei and Varga}{2023}]{szendrei2023revisiting}
Szendrei, T. and K.~Varga (2023).
\newblock Revisiting vulnerable growth in the euro area: Identifying the role of financial conditions in the distribution.
\newblock {\em Economics Letters\/}~{\em 223}, 110990.

\bibitem[\protect\citeauthoryear{Wolters and Tillmann}{Wolters and Tillmann}{2015}]{wolters2015changing}
Wolters, M.~H. and P.~Tillmann (2015).
\newblock The changing dynamics of us inflation persistence: A quantile regression approach.
\newblock {\em Studies in Nonlinear Dynamics \& Econometrics\/}~{\em 19\/}(2), 161--182.

\bibitem[\protect\citeauthoryear{Yu and Jones}{Yu and Jones}{1998}]{yu1998local}
Yu, K. and M.~Jones (1998).
\newblock Local linear quantile regression.
\newblock {\em Journal of the American statistical Association\/}~{\em 93\/}(441), 228--237.

\bibitem[\protect\citeauthoryear{Zou and Yuan}{Zou and Yuan}{2008}]{zou2008composite}
Zou, H. and M.~Yuan (2008).
\newblock {Composite quantile regression and the oracle model selection theory}.
\newblock {\em The Annals of Statistics\/}~{\em 36\/}(3), 1108 -- 1126.

\end{thebibliography}

\pagebreak


\appendix 
\section{Appendix}
\counterwithin{myfigure}{section}
\counterwithin{table}{section}
\setcounter{table}{0}
\setcounter{figure}{0}

\begin{myfigure}[h]
    \centering
    \includegraphics[width=0.875\textwidth]{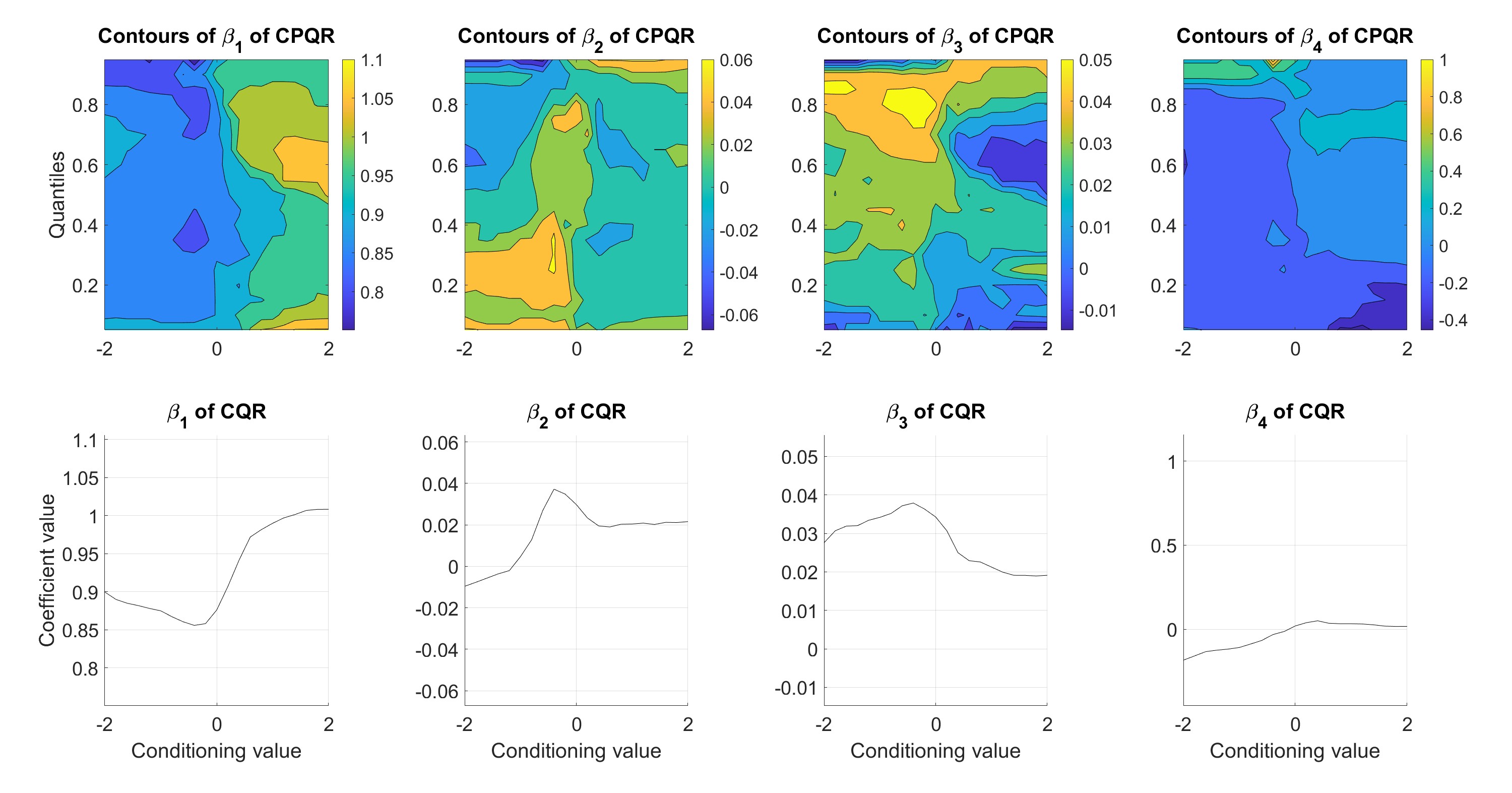}
    \caption{Coefficients of CPQR without constraints for h=1}
    \label{fig:CPQRnoNC_h1}
\end{myfigure}

\begin{myfigure}[h]
    \centering
    \includegraphics[width=0.875\textwidth]{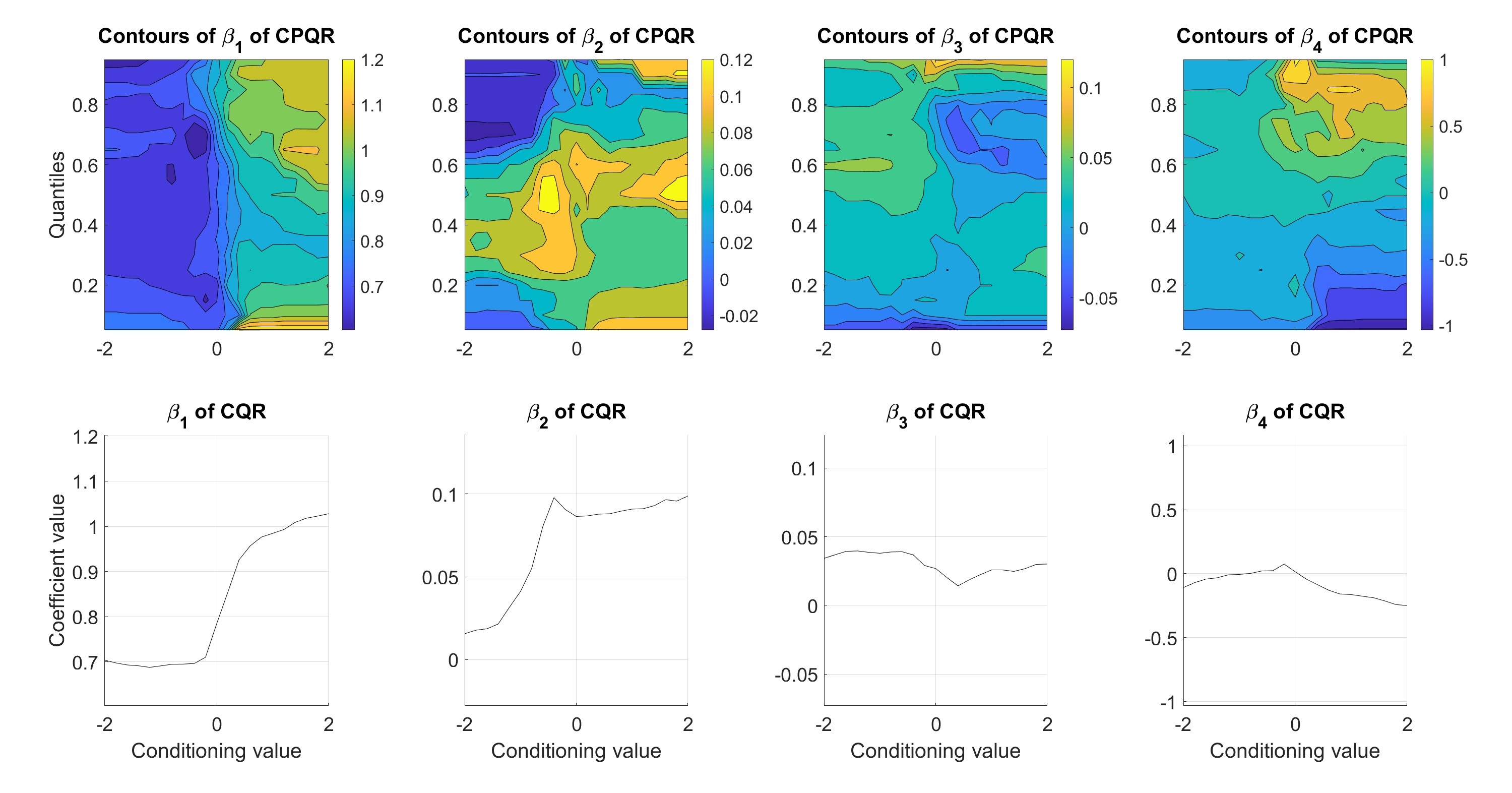}
    \caption{Coefficients of CPQR without constraints for h=4}
    \label{fig:CPQRnoNC_h4}
\end{myfigure}

\end{document}